%% file: ml2020.tex
\documentclass[letterpaper,twocolumn,10pt]{article}
\usepackage{./usenix2019_v3}
\usepackage[pass]{geometry}

\usepackage[english]{babel}
\usepackage{csquotes}
\usepackage{array}
\usepackage{textcomp}
\usepackage{todonotes}
\usepackage{graphicx}
\usepackage{float}

\usepackage{url}

\usepackage[style=numeric,
			sorting=none,
			giveninits=true,
			maxnames=99,
			isbn=false]{biblatex}
\usepackage{subfig}

\usepackage{xspace}
\usepackage{siunitx}
\DeclareSIQualifier\peak{p}
\DeclareSIQualifier\peakpeak{pp}
\DeclareSIUnit{\pixel}{px}
\newcommand{\sixyfourtimespx}{64\,\si{\pixel}\,$\times$\,64\,\si{\pixel}\xspace}
\newcommand{\hundredtwentyeighttimespx}{128\,\si{\pixel}\,$\times$\,128\,\si{\pixel}\xspace}

\usepackage{layouts}
\usepackage{placeins}

\usepackage{enumitem}

\usepackage{import}

\usepackage{tikz}
\usepackage{pgfplots}
\usepgfplotslibrary{groupplots,dateplot}
\usetikzlibrary{patterns,shapes.arrows}
\pgfplotsset{compat=newest}

\newsavebox\mybox
\newlength{\axwidth}	
\newlength{\axheight}
\definecolor{color0}{RGB}{161,218,180}
\definecolor{color1}{RGB}{65,182,196}
\definecolor{color2}{RGB}{34,94,168}
\tikzset{MarkStyle/.style={solid, draw=black, thin}}
\tikzset{BgStyle/.style={fill=white!95!black}}

\usepackage{caption}

\usepackage{setspace}

\DeclareRobustCommand*{\authorrefmark}[1]{\raisebox{0pt}[0pt][0pt]{\textsuperscript{\footnotesize\ensuremath{\ifcase#1\or *\or \dagger\or \ddagger\or%
	\mathsection\or \mathparagraph\or \|\or **\or \dagger\dagger%
	\or \ddagger\ddagger \else\textsuperscript{\expandafter\romannumeral#1}\fi}}}}

\usepackage{ifthen}
\newboolean{cameraready}
\setboolean{cameraready}{true}

\newboolean{arxiv}
\setboolean{arxiv}{true}

\ifthenelse{\boolean{arxiv}}{
\usepackage[printwatermark]{xwatermark}
\newwatermark*[firstpage,color=black!70,angle=0,xpos=-55,ypos=-125, align=center, fontsize=6.8,fontseries=m]{%
	\xwmminipage[width=.25\linewidth,textalign=justified]{\textit{Note:} This is the preprint of the article accepted for publication at USENIX Security 2021.}}
}{}

\usepackage{hyperref}

\begin{document}

\ifthenelse{\boolean{cameraready}}{
	\ifthenelse{\boolean{arxiv}}{
	}{
	\pagestyle{empty}
	}
}{}

\date{}
	
\title{
Automatic Extraction of Secrets from the Transistor Jungle using\\ Laser-Assisted Side-Channel Attacks
}

\ifthenelse{\boolean{cameraready}}{
\author{
	{\rm Thilo Krachenfels\authorrefmark{1}, Tuba Kiyan\authorrefmark{1}, Shahin Tajik\authorrefmark{2} and Jean-Pierre Seifert\authorrefmark{1}\authorrefmark{3}}\\
	\authorrefmark{1} Technische Universit\"at Berlin, Chair of Security in Telecommunications\\
	\authorrefmark{2} Worcester Polytechnic Institute, Department of Electrical and Computer Engineering\\
	\authorrefmark{3} Fraunhofer SIT
} %
}{
\author{
	{\rm Anonymous}
} %
}

\maketitle

\section*{Abstract}

The security of modern electronic devices relies on secret keys stored on secure hardware modules as the root-of-trust (RoT).
Extracting those keys would break the security of the entire system.
As shown before, sophisticated side-channel analysis (SCA) attacks, using chip failure analysis (FA) techniques, can extract data from on-chip memory cells.
However, since the chip's layout is unknown to the adversary in practice, secret key localization and reverse engineering are onerous tasks.
Consequently, hardware vendors commonly believe that the ever-growing physical complexity of the integrated circuit (IC) designs can be a natural barrier against potential adversaries.
In this work, we present a novel approach that can extract the secret key without any knowledge of the IC's layout, and independent from the employed memory technology as key storage.
We automate the – traditionally very labor-intensive – reverse engineering and data extraction process.
To that end, we demonstrate that black-box measurements captured using laser-assisted SCA techniques from a training device with known key can be used to profile the device for a later key prediction on other victim devices with unknown keys.
To showcase the potential of our approach, we target keys on three different hardware platforms, which are utilized as RoT in different products.

\section{Introduction}

For security applications, people rely on hardened hardware modules, like Trusted Platform Modules (TPMs), as the root-of-trust (RoT) for storing secret keys.
Those keys ensure the functioning of complex and delicate systems like routers, servers, sensor systems, and cars by establishing secure communication channels, safeguarding trusted code execution, and protecting the intellectual property embodied in the device.
Extracting secret keys managed by a RoT hardware would break the entire system's security.
Possible motivations for attackers are the extraction of secret information, tampering with the design, or cloning the device.

Modern integrated circuits (ICs) and system-on-chips (SoCs) consist of billions of transistors, which makes the reverse engineering of the design and layout very challenging.
Moreover, data extraction from various key storage technologies requires different measurement tools and expertise, making the attack costly and unscalable.
This physical complexity might lead to a belief by vendors that the localization and extraction of assets/secrets on their products is a laborious task.
In addition to that, the usage of the keys in diverse applications, such as firmware/bitstream decryption, asymmetric cryptographic operations, or logic deobfuscation, makes the generalization of an attack against RoTs infeasible.

There are companies like Techinsights~\cite{techinsigh_semiconductor_2020} and Texplained~\cite{texplained_2021} that invest lots of expertise and effort into fully reverse engineering ICs with destructive techniques and using sophisticated failure analysis (FA) tools, such as scanning electron microscopes (SEMs) and focused ion beams (FIBs)~\cite{quadir_survey_2016}.
They can extract the IC's netlist, analyze its functioning, and therefore, find the location where the key is stored.
While effective, this approach is very time consuming and expensive.
On the other hand, researchers have shown that attacks on some specific devices only require partial reverse engineering.
Applying SEM~\cite{kison_finding_2015, courbon_reverse_2017}, FIB~\cite{helfmeier_breaking_2013}, microprobing~\cite{kommerling_design_1999}, and laser-assisted side-channel analysis (SCA) techniques using laser scanning microscopes (LSMs) ~\cite{lohrke_no_2016,tajik_power_2017,lohrke_key_2018,krachenfel_realworld_2020} are examples of such academic work.
Nevertheless, these attacks have only been carried out in an experimental environment, where many details of the design were available beforehand or had to be gathered manually.

Considering the high amount of manual reverse engineering work, one might ask if machine learning techniques could be applied in the context of hardware security to reduce the required knowledge for key extraction. %
Indeed, the benefit of applying deep learning techniques on classical SCA attacks, like power and electromagnetic (EM) analysis, have already been discovered and studied extensively~\cite{maghrebi_breaking_2016,benadjila_study_2018,kubota_deep_2019,hou_convolutional_2019}.
At the same time, convolutional neural networks (CNNs) have become the default choice for image classification tasks, as they remove the need for manually tailoring the algorithm to its specific application.
Consequently, CNNs could also be one suitable method for extracting a key from images captured by FA techniques from a complex chip with unknown design and layout.
In other words, if an attacker combines image recognition techniques with sophisticated FA tools that are capable of capturing the logic state from inside the IC, a new threat dimension arises.
Such an approach can antiquate the expensive reverse engineering portion of hardware attacks.
On the positive side, such a tool, if automated, can also be used for security assessment of products. 

\vspace{.5mm}
\noindent
\textbf{Our contribution.} 

In this work, we develop an attack approach drawing the connection between image recognition techniques, profiling SCA, and sophisticated FA tools to extract the secret key from memory cells of an IC without requiring any knowledge about the chip's layout and its functioning.

To validate our claims, we conduct SCA using two different and well-known laser-assisted SCA methods, namely thermal laser stimulation (TLS)~\cite{lohrke_key_2018} and laser logic state imaging (LLSI)~\cite{krachenfel_realworld_2020}.
We apply these SCA techniques on three different hardware targets with various process technology sizes: the dedicated key memory of an 20\,nm Field Programmable Gate Array (FPGA), the SRAM of a 180\,nm microcontroller, and the registers of an 60\,nm FPGA.
All these platforms can be potentially part of an RoT implementation.
To showcase the strength of our approach, we exemplarily deploy CNNs to create models out of obtained measurements from these devices.
The results demonstrate that our trained models can extract an unknown secret key from the victim devices with high accuracy, even in the presence of largely irrelevant information and activities on the chip.
Moreover, it is not required to know the location of targeted memory cells and how to interpret the bit values from the measurements.
Note that our approach is not limited to optical SCA attacks, and can also be combined with SEM, FIB, or any other FA microscopy tools, which capture the activity of transistors.

While in this work we have applied deep learning due to its straight-forward nature for highlighting the threat of our approach, deploying other image recognition techniques is also conceivable.
In this regard, we are open-sourcing the side-channel data to enable other researchers to improve data extraction using various techniques.
Consequently, we would like to stress that the emphasis of this work is on showing that laser-based SCA can eliminate the reverse engineering step for extracting secret information, and not on applying deep learning techniques as profiling SCA tool.

\section{Threat Model\label{sec:Threat-Model}}

\subsection{Target}

\begin{figure}
	\centering
	\includegraphics[width=\linewidth]{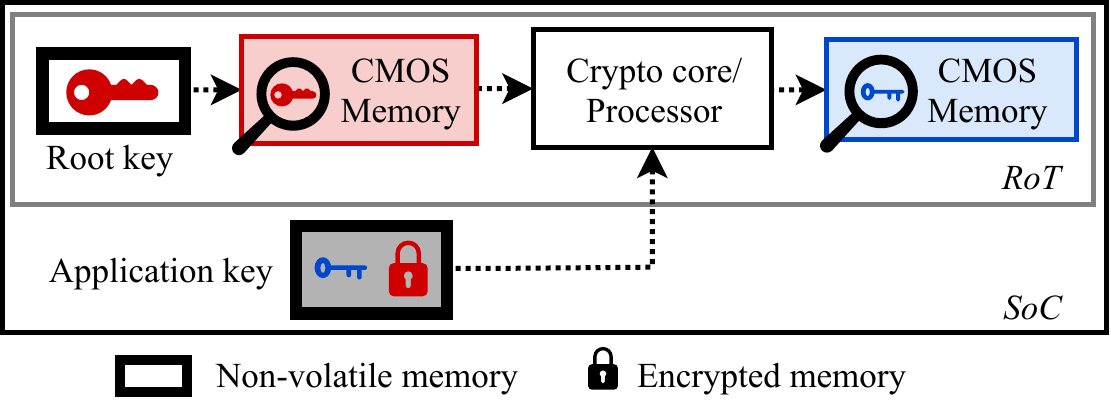}
	\vspace{-5mm}
	\caption{Extraction of the root key after it was loaded from a tamper- and read-proof non-volatile memory, or of an application key after it has been decrypted using the root key.\label{fig:threat_model}} %
\end{figure}

In hardware RoT applications, we can distinguish between two different kinds of keys.
At least one \textit{root key} must be stored in plaintext in a non-volatile memory (NVM). 
Other keys might be stored internally/externally in an encrypted form, decryptable by the root key.
In the following, we will refer to them as \textit{application keys}.
In addition to its usage as key-decryption-key, the root key might also be deployed directly, e.g., for firmware or bitstream decryption.
Since the root key is typically stored in a (tamper- and read-proof) NVM, such as flash memory, EEPROM, or ferroelectric RAM, the direct extraction of its content is not a straightforward task~\cite{courbon_reverse_2017}.
However, even if the NVM is considered secure, for being used, the contained key will be loaded into CMOS memory cells at some point in time, see Fig.~\ref{fig:threat_model}.
The same holds true for application keys after they have been decrypted.

Previous work has shown that sophisticated non- and semi-invasive FA tools are capable of extracting logic states from CMOS logic gates~\cite{rahman_key_2020} and memory cells~\cite{lohrke_no_2016,lohrke_key_2018,krachenfel_realworld_2020}.
These techniques typically produce an image (i.e., activity map or response image) which contains information about the logic state of the area of interest.
Yet, extracting the actual memory content from these images can be a challenging task, even if the chip's layout is known, or at least understood to a certain degree.
Although tools like SAT solvers~\cite{krachenfel_realworld_2020} and image recognition techniques can aid the localization of the key, much prior knowledge of the memory cell's design, its geometry, and its exact location is required.
Therefore, a potential attacker might be highly motivated to reduce the effort for extracting keys from the images.

\subsection{Attacker's Motivation}

We assume an adversary who has access to FA tools and has a strong motivation to avoid expensive reverse engineering of the whole IC for just extracting a single key out of it.
One might ask why an adversary would invest that much effort into extracting the key from a single device.
The primary motivation in many scenarios is that the same key is used for all devices, for instance, when firmware, bitstream, or logic encryption is used to protect the proprietary design of a system.
The key is therefore programmed by the vendor before the product is shipped to the customer.
Consequently, extracting the key from one device would break the security of all devices from the same family.
Even if the key differs between devices, it should be kept in mind that all chips from a device family have the same layout.
Therefore, the adversary can learn how to extract the key from a training device and use her knowledge to extract the key also from other devices of the same family.

\section{Background\label{sec:Background}}

\subsection{Optical Side-Channel Analysis Attacks\label{subsec:Optical-Side-Channel-Analysis}}

For being able to debug the active silicon of integrated circuits (ICs) in the presence of the many metal layers on the chip frontside, techniques have been developed to access on-chip signals through the IC backside~\cite{boit_ic_2016}.
The corresponding optical side-channel analysis (SCA) techniques take advantage of the high infrared transmission in silicon for wavelengths above 1\,\si{\micro\meter}, basically allowing to ``see through'' the bulk silicon at the IC backside.
Due to their availability in FA labs around the globe, related techniques like photon emission analysis, laser stimulation, and optical probing have been adopted by the hardware security field~\cite{tajik_emission_2014,lohrke_key_2018,tajik_power_2017}.
A typical setup consists of a laser scanning microscope (LSM) with laser sources of different wavelengths, a detector for measuring the reflected laser light, and optionally a camera for photon emission analysis.

The two relevant techniques for this work, including reported attacks in the literature, will be discussed below.

\subsubsection{Thermal Laser Stimulation\label{subsec:Thermal-Laser-Stimulation}}

\begin{figure}
\centering{}\subfloat[TLS\label{fig:TLS_schematic}]{
{\footnotesize
	\def\svgwidth{.48\linewidth}
	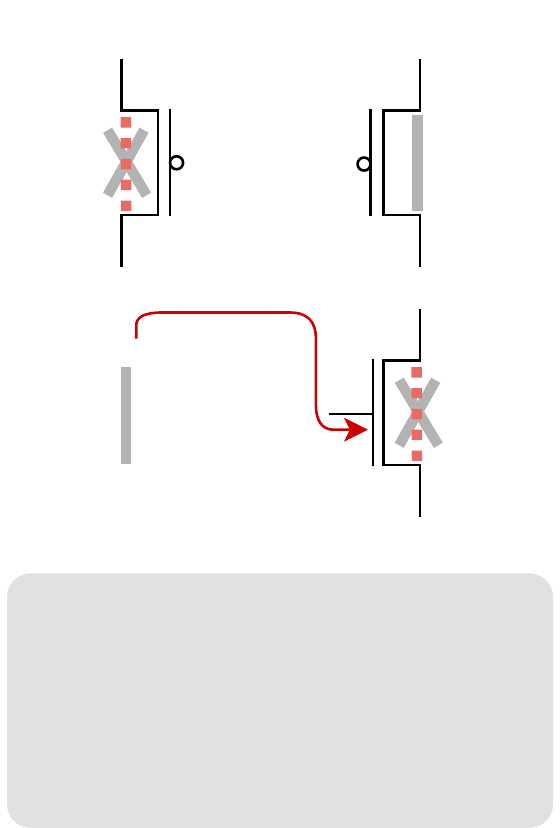
	\vspace{1mm}
}

}\subfloat[LLSI\label{fig:LLSI_schematic}]{
{\footnotesize
	\def\svgwidth{.48\linewidth}
	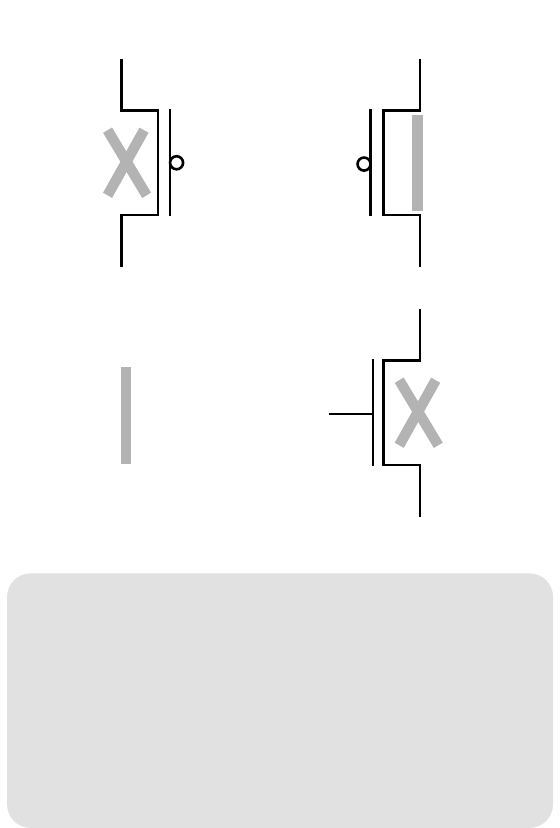	
	\vspace{1mm}
}}
\vspace{-2mm}
\caption{Schematic of a CMOS memory cell and how the two measurement techniques can extract the cell's logic state. Transistors for read and write access are omitted. Figures based on~\cite{nedospasov_invasive_2013,niu_laser_2014}.\label{fig:TLS-LLSI-Schematic}}
\end{figure}

Thermal laser stimulation (TLS) is an SCA technique that induces electrical perturbations on a target device by creating local temperature gradients when stimulating an area of interest with a laser beam.
The laser beam's wavelength is above 1.1\,\si{\micro\meter}, which does not have enough energy to generate electron-hole pairs, but thermal gradients.
A temperature variation on a thermocouple can lead to a voltage generation, which is known as the Seebeck effect~\cite{geballe_seebeck_1955}.
The Seebeck voltage can be leveraged to extract the logical states from CMOS memory cells~\cite{nedospasov_invasive_2013}.

A CMOS memory cell consists of two cross-coupled inverters, with one transistor per inverter being low-ohmic (conducting) and one being high-ohmic (nonconducting), see Fig.~\ref{fig:TLS_schematic}.
Hence, while storing a value, i.e., in the stable state, only a negligible current is flowing between VCC and GND. However, if a laser beam stimulates the drain-bulk junction of a transistor with low-ohmic channel, it generates a Seebeck voltage (V\textsubscript{Seeb.}).
This voltage is forwarded along the circuit to the gate of a transistor in the high-ohmic state.
This transistor is slightly switched on and -- via exponential sub-threshold operation--, the current drawn from the memory cell's power supply increases.
If an area of interest on the device is scanned pixel-wise by a laser beam and the small power consumption variations are recorded along with the laser beam's location, the TLS response map of the scanned area can be obtained.
The areas of the two sensitive transistors will show up brighter in the TLS response map, due to the slight increase in power consumption.
For the opposite bit state, the other two transistors will appear on the response map, making the two different bit states of the memory cell distinguishable from each other.

TLS is a well-understood technique that has been used to read out SRAM memory on microcontrollers~\cite{nedospasov_invasive_2013,kiyan_comparative_2018} and extract the cryptographic key from the battery-backed RAM on an FPGA~\cite{lohrke_key_2018}.
One scan over the area of interest can reveal the entire memory content, and therefore, TLS can be considered a single-trace SCA technique.
Naturally, the memory content should stay constant during the scan.
Recently it has been shown that TLS can be mounted with cheaper setups than previously
expected -- for around \$100k~\cite{krachenfel_evaluation_2020}.

\subsubsection{Laser Logic State Imaging\label{subsec:Optical-Probing}}

Optical probing is an FA tool used for acquiring electrical information from inside the IC~\cite{yee_laser_1999,lohrke_no_2016,tajik_power_2017}. %
Electro-optical frequency mapping (EOFM) is an optical probing technique that allows creating a 2-D activity map of circuits, showing nodes that are switching at a particular frequency~\cite{zhang_electro_2017}.
While light with wavelengths above 1~\si{\micro\meter} scans the IC backside pixel by pixel, it passes through the silicon substrate.
The light is partially absorbed and partially reflected by structures such as metal layers and transistors, whereas the electrical field present at transistors influences the light's amplitude and phase.
A portion of the reflected light leaves the IC through the backside where it is converted into a voltage and fed into a narrow-band frequency filter set to the frequency of interest.
The resulting signal's amplitude and the position information form the 2-D activity map on which areas modulating at the frequency of interest appear as bright spots.

For EOFM measurements, it is necessary to know the internal switching frequency of the circuit of interest to track the signals.
This frequency can be hard to predict, and even worse, there is not necessarily any switching activity for memory cells if no read/write operation is carried out.
This problem can be tackled by inducing a frequency, for example, by modulating the core voltage that supplies the circuit under test.
The corresponding technique is called laser logic state imaging (LLSI) and has been introduced as an extension to EOFM~\cite{niu_laser_2014}.
LLSI makes the extraction of static logic states possible, e.g., from a CMOS memory cell, as illustrated in Fig.~\ref{fig:LLSI_schematic}.
The low-ohmic transistors' electric fields oscillate with the power supply's modulation frequency, and hence, produce an EOFM signal.
In contrast, off-state transistors do not produce a strong EOFM signal.
Consequently, the logic state of the SRAM cell can be deduced.
LLSI has been used to read out SRAM on a microcontroller~\cite{kiyan_comparative_2018} and the registers on an FPGA~\cite{krachenfel_realworld_2020}.

Note that LLSI can be used to extract not only the state of SRAM cells or registers, but also any cluster of transistors, such as buffers or logic gates.
As long as the bit state of the logical element affects the involved transistors, the bit value can be extracted.
Next to TLS, also LLSI can be considered a single-trace SCA technique, as one scan over the region of interest is sufficient to capture its entire logic state.
Similarly, to perform LLSI, the memory content has to remain constant during the scan.
One way to achieve this requirement is to halt the clock signal to prevent any update in the values of the memory~\cite{krachenfel_realworld_2020}.
However, in some applications, e.g., logic locking, the secret key has to be provided constantly to the locked circuit in order to keep it unlocked during runtime, and therefore, no clock control is needed.
Moreover, it has been observed that some cryptographic accelerators do not necessarily clear key registers after encryption/decryption~\cite{moos_static_2019}, and hence, the key remains in the registers as long as the device is powered on.

\subsection{Deep Learning for Image Classification\label{subsec:Background_Deep-Learning}}

Due to their high flexibility, convolutional neural networks (CNNs)~\cite{lecun_backpropagation_1989,lecun_gradientbased_1998,lecun_learning_2004} are a popular choice for many computer vision applications such as image recognition~\cite{russakovsk_imagenet_2015,liu_deep_2020}.
Image recognition typically consists of two tasks: object classification (also called image-level annotation) and object detection (object-level annotation).
While for classification only the presence of an object from a given set of classes is assessed -- and not its position --, object detection is typically a more challenging task.
In this work, we are only interested in the existence of a logic \texttt{0} or \texttt{1} in an image, and therefore, we will only cover object classification in the following.

CNNs are a subclass of deep neural networks and complement the fully-connected (FC) networks (also known as multilayer perceptrons) with trainable feature extractors, the so-called \textit{convolutional layers}.
A convolutional layer finds features in the image (e.g., corners, edges, etc.) using trainable filters that cover a certain receptive field.
The resulting \textit{feature maps} can be fed into subsequent convolutional layers to detect larger features. Intermediate subsampling steps -- \textit{pooling layers} -- reduce the resolution of the feature maps to decrease the sensitivity to shifts and other distortions.
Finally, after some repetitions of convolutional and pooling layers, the output is \textit{flattened} and fed into the FC network to classify the images.

In the literature, different architectural designs for CNNs have been reported, e.g., LeNet-5~\cite{lecun_gradientbased_1998}, AlexNet~\cite{krizhevsky_imagenet_2012}, and VGG~\cite{simonyan_very_2015}.
The authors of the VGG architecture presented a generic design consisting of the repetitive application of filters with a very small receptive field (3$\times$3 pixels), followed by a max-pooling over a 2$\times$2 pixel window.
The stack of convolutional layers is followed by FC layers with one neuron for each class in the output
layer \cite{simonyan_very_2015}.
The structure of multiple small convolutional layers followed by a max-pooling layer is often referred to as \textit{VGG-block} and has become a popular building block and starting point when designing a new model from scratch, like it will be required for the optical key extraction.
Different concepts have been developed to reduce over-specialization on the training data (so-called overfitting) of CNNs, especially when only a small training dataset is available.
For instance, a \textit{dropout layer} can remove random nodes from the FC layers during training, which leads to the extraction of more robust features~\cite{hinton_improving_2012}.
Furthermore, \textit{data augmentation} can increase the number of training samples artificially, and therefore, reduce overfitting as well~\cite{simard_best_2003}.

\begin{figure*}[ht!]
	\centering
	\includegraphics[width=.9\linewidth]{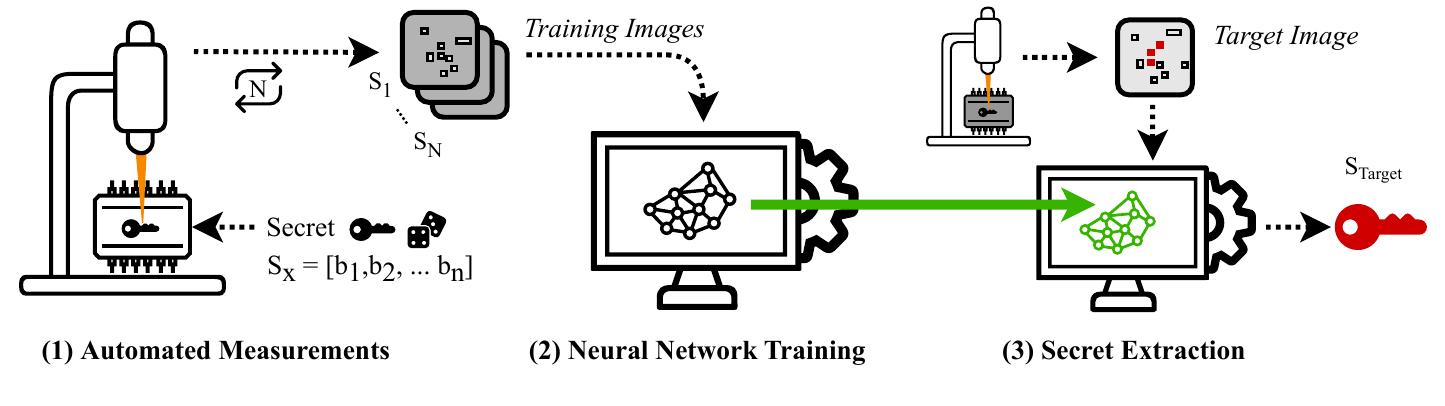}
	\vspace{-6mm}
	\caption{Schematic of the proposed three-step attack approach.
		\label{fig:Approach}}
\end{figure*}

\subsection{Related Work\label{subsec:Related-Work}}

This work builds on an approach that is known as profiled side-channel analysis~\cite{lerman_template_2015}, where a device under the adversary's control is used to create a leakage model, which is later used to extract the secret from a similar device~\cite{standaert_how_2009}.
In the literature, profiled SCA is typically applied to a cryptographic core by observing its operation, for instance, through power and EM side-channels.
In the \textit{profiling phase}, the behavior of the DUT is observed and incorporated in a leakage model using either statistical methods (a.k.a template attacks~\cite{choudary_template_2014}) or machine learning techniques~\cite{elnaggar_machine_2018}, such as support vector machines~\cite{hospodar_machine_2011} and neural networks~\cite{maghrebi_breaking_2016, benadjila_study_2018, kubota_deep_2019, hou_convolutional_2019}.
In the \textit{attack phase}, the extracted model is used to extract the unknown secret from the target device.
Traditional SCA has limited applicability in some cases, e.g., when the key is not involved in active computations, or when countermeasures prohibit the capturing of a sufficient number of traces.

Next to side-channel analysis, machine learning is also used in many other applications in the field of hardware security~\cite{elnaggar_machine_2018}, for instance, for hardware trojan detection~\cite{hasegawa_hardware_2017} and reverse engineering~\cite{quadir_survey_2016, chen_deep_2020}.

\section{Attack Approach\label{sec:Scenario-and-Approach}}

Our attack approach%
\ifthenelse{\boolean{cameraready}}{
has already been sketched in~\cite{boit_logic_2020} and%
}{}
assumes that the adversary has access to a training device, for which she can control the contained secret at her will%
.
However, she does not have any knowledge about the design of the chip and the location of the key storage. 
In this scenario, the approach for the attacker consists of three steps, see Fig.~\ref{fig:Approach}.
In the first step, randomly chosen keys are programmed into the training device, and SCA images are captured from the IC backside for each key.
Subsequently, neural networks are trained with the obtained images.
These two steps can be specified as profiling phase.
In the final step, the attack phase, the secret on the target device is revealed by one or a few measurements and the previously trained networks.
Note that in this work, we chose to apply deep learning techniques for image recognition due to their ad-hoc adaptability to many problems with minimal tuning effort.
For the secret extraction from the images, potentially also other machine learning or statistical methods can be applied.
In the following, we discuss the three steps of our approach in more detail.

\subsection{Automated Measurements\label{subsec:Automated-Measurements}}

For gathering a training dataset, the adversary captures response images using TLS or LLSI from the training device containing different randomly chosen keys.
Since capturing many high-resolution images from larger areas of the chip can be very time consuming, the attacker would first try to find candidate areas for the on-chip memory.
Due to the repetitive and regular structure of memory arrays, such candidate areas often can be discovered by analyzing an optical image of the chip.
If this is not the case, two response images (containing two different secrets) can be captured from the entire chip area.
When subtracting the two images, the attacker can consider all areas showing a difference as candidate areas which should be covered by the automated measurements.
Consequently, one sample in the training database consists of one or more response images and the programmed secret.
After capturing some samples, the attacker can continue with step~2, that is, training CNNs with the database.

\subsection{Neural Network Training\label{subsec:Neural-Network-Training}}

Before training CNNs with the response images, possible drift caused by mechanical instabilities of the setup should be corrected.
For this, classical image registration techniques can be used, e.g., by calculating the offset between an optical image captured along with the response image and one fixed optical image.
Subsequently, the response image can be transformed according to the calculated shift.

Furthermore, the programmed secret is split into its individual binary bits, which are assigned as multiple labels to each image -- one label per bit.
Once these preparatory steps are done, a CNN can be designed to learn the secret bits from the response images.
More specifically, for each bit of the secret, the images are classified to contain either the binary bit value \texttt{0} or \texttt{1}.
Note that each bit of the secret is handled independently from the other bits.
To find out if the images depend on the secret at all, different network architectures should be investigated while trying to learn just a single bit of the secret.
Following common practice, we propose to start with a simple model, containing only a few convolutional layers (one VGG-block, see Section~\ref{subsec:Background_Deep-Learning}).

To reduce the resources needed for training the model, the images can be split into smaller-sized sections, and a separate model can be trained on each section.
As a side-effect, the attacker can find the secret's rough location.
If the network does not reach a very high validation accuracy, but the secret bits can be learned to some degree, more measurements from the respective section might be required (supposedly also with higher resolution).
The application of data augmentation techniques is likely to reduce the required number of measurements.
Once single bits can be learned successfully, a multi-label classification can be attempted to reduce the training time.
In other words, one network should learn more than one bit at the same time.
This can be achieved by adding more output neurons to the FC network -- one per bit of the key to be learned.

\subsection{Secret Extraction\label{subsec:Secret-Extraction}}

When all bits could be learned using the training dataset with a sufficiently high accuracy, the attacker knows the required locations on the chip and measurement parameters for a successful extraction of the secret.
She then can capture response images from the target device (containing an unknown secret) and let the obtained models predict the key from those images.
Depending on the accuracy of the network, multiple images with slightly different parameters (like focus position) could be obtained for being able to apply a majority voting scheme on the predicted secret bits, and therefore, achieve a higher probability for predicting all bits of the secret correctly.
In this work, we abstain from extracting the secret from a target device and instead rely on the test accuracy from the training phase as an indicator for the attack's success.
However, we expect the inter-device differences to be lower than the noise introduced during different measurement runs and by data augmentation.

\section{Experimental Setup and Target Devices\label{sec:Setup-and-Target}}

In the first part of this section we give details on our setup for conducting TLS and LLSI measurements%
.
Then we briefly describe our setup for the learning part. Finally, we introduce the devices under test (DUTs) and present images of their memory structures captured with our setup.

\subsection{Measurement Setup}

\subsubsection{Optical and Electrical Setup}

The core of our setup is a Hamamatsu PHEMOS-1000 FA microscope. It is equipped with a 1.3\,\si{\micro\meter} high-power incoherent light source (HIL) for optical probing and a 1.3\,\si{\micro\meter} laser for thermal stimulation.
In addition to the 5$\times$, 20$\times$, and 50$\times$ lenses, a scanner-zoom of 2$\times$, 4$\times$, and 8$\times$ is available.
The light beam is scanned pixel-wise over the device using galvanometric mirrors.
For acquiring optical images and conducting LLSI, the reflected light is separated by semi-transparent mirrors and fed into a detector.
For LLSI, the detector's output is fed into a bandpass filter set to the frequency of interest.
The PC software then produces a 2-D image containing the measured amplitude at each pixel.
For conducting TLS measurements, the laser is scanned over the device, and its power consumption is measured using an external current preamplifier (Stanford Research Systems SR 570).
The amplifier's output is fed into the \mbox{PHEMOS} PC software, which produces a response map of the locations sensitive to the thermal stimulation.
The setup specific to the devices under test is described in Section~\ref{subsec:Devices-under-Test}.

\subsubsection{Measurement Automation}

For repeating the measurements with different secrets programmed into the target devices, we programmed a tool in the LabView programming environment.
It can control the \mbox{PHEMOS} software (e.g., start and stop measurements, execute auto-focus, move the lens) and access the captured images for correcting horizontal and vertical drift.
Furthermore, the tool can trigger the programming of a new secret into the DUT by communicating with a target-specific script running on another PC.
In one iteration of the automated measurements, first a new secret is programmed.
Then, after executing the auto-focus, an optical image is captured and saved.
The drift between that image and the first image of the measurement series is calculated and the lens is moved accordingly.
Finally, the TLS or LLSI measurement is conducted and the resulting image is saved along with the secret.

\subsection{Learning Setup}

For correcting drift in the final images, we made use of the MATLAB image processing toolbox.
As machine learning toolbox, we used the Keras API for TensorFlow (version 2.3.0).
We ran all our experiments on an Ubuntu 20.04.1 LTS machine with an Intel i7-6850K CPU~@\,3.6\,GHz, 128\,GiB of system memory and a GeForce GTX 1080 Ti GPU with 11\,GiB of memory.
For all experiments, we made use of the TensorFlow GPU support.

\subsection{Devices under Test\label{subsec:Devices-under-Test}}

We chose three different targets manufactured in different technology sizes and containing different kinds of volatile key memories for our evaluations.
\begin{figure}[t]
	\centering
	\subfloat[Optical image]{\includegraphics[width=0.85\linewidth]{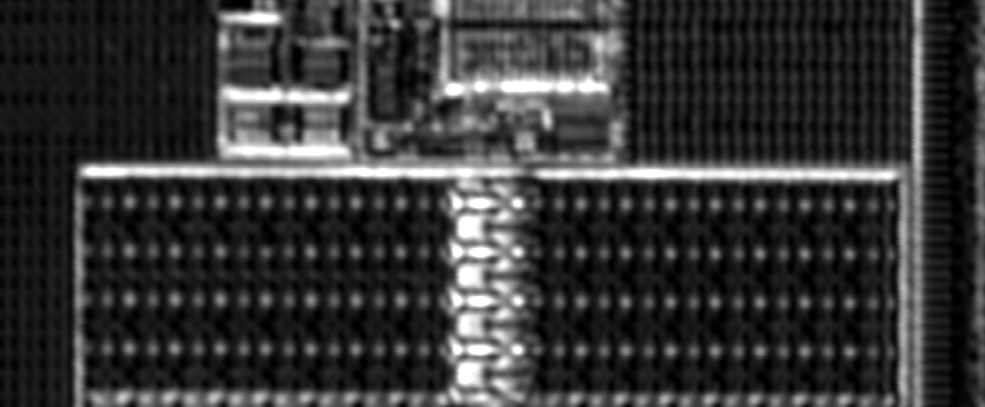}}
	\vspace{-2mm}
	\subfloat[TLS response image with a random key programmed]{\includegraphics[width=0.85\linewidth]{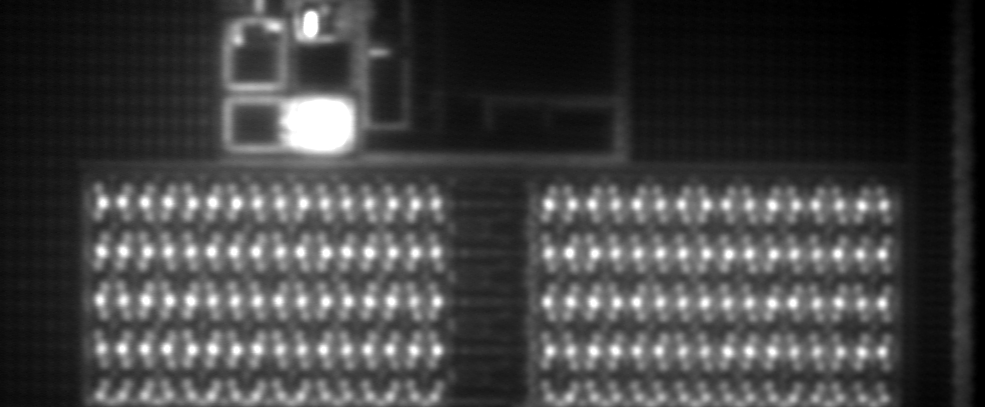}}
	\vspace{-2mm}
	\subfloat[Difference between two TLS response images with different keys\label{subfig:BBRAM-target-images_diff}]{\includegraphics[width=0.85\linewidth]{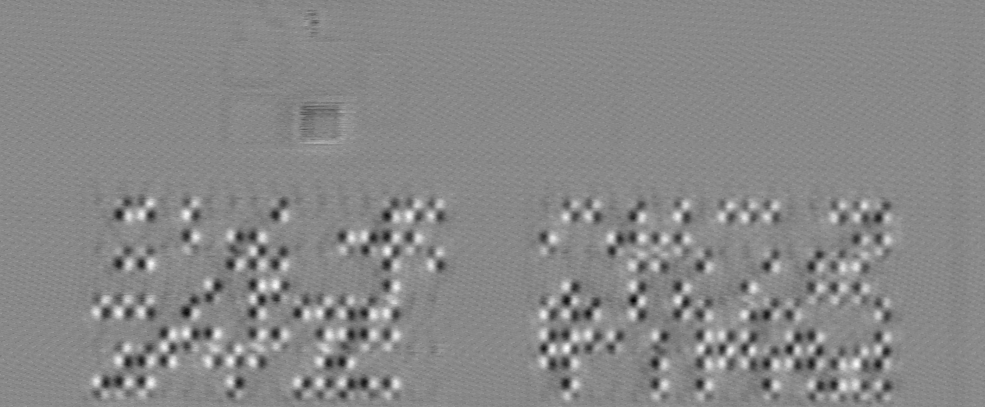}}
	\vspace{-1mm}
	\caption{Images of the Xilinx Ultrascale BBRAM.\label{fig:BBRAM-target-images}\vspace{-4mm}}
\end{figure}

\begin{figure*}[htb]
	\subfloat[Optical image]{\includegraphics[width=0.21\textwidth]{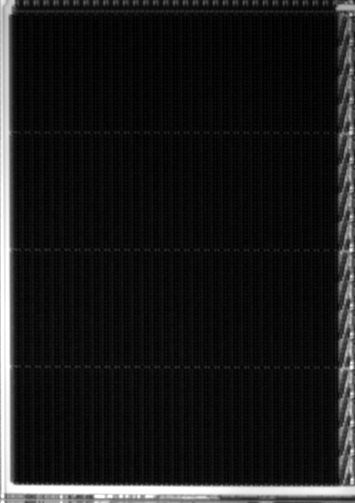}}
	\hfill
	\subfloat[LLSI image (512 key bits, rest zeroized)]{\includegraphics[width=0.21\textwidth]{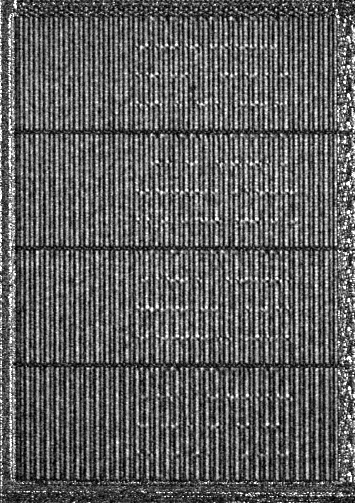}}
	\hfill
	\subfloat[LLSI image (512 key bits, rest randomized)]{\includegraphics[width=0.21\textwidth]{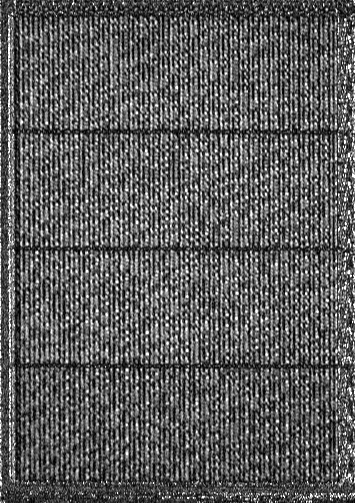}}
	\hfill
	\subfloat[Difference between two LLSI images (rest zeroized)\label{subfig:MSP430-target-images_diff}]{\includegraphics[width=0.21\textwidth]{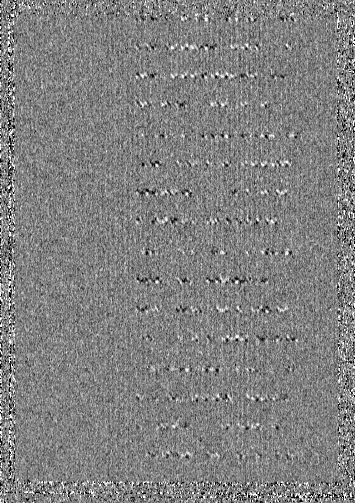}}
	\vspace{-1mm}
	\caption{Images of the TI MSP430's 1024-byte SRAM area.\label{fig:MSP430-target-images}}
\end{figure*}

\begin{figure}[htb]
	\centering
	\subfloat[Optical image]{\includegraphics[angle=90,width=0.45\textwidth]{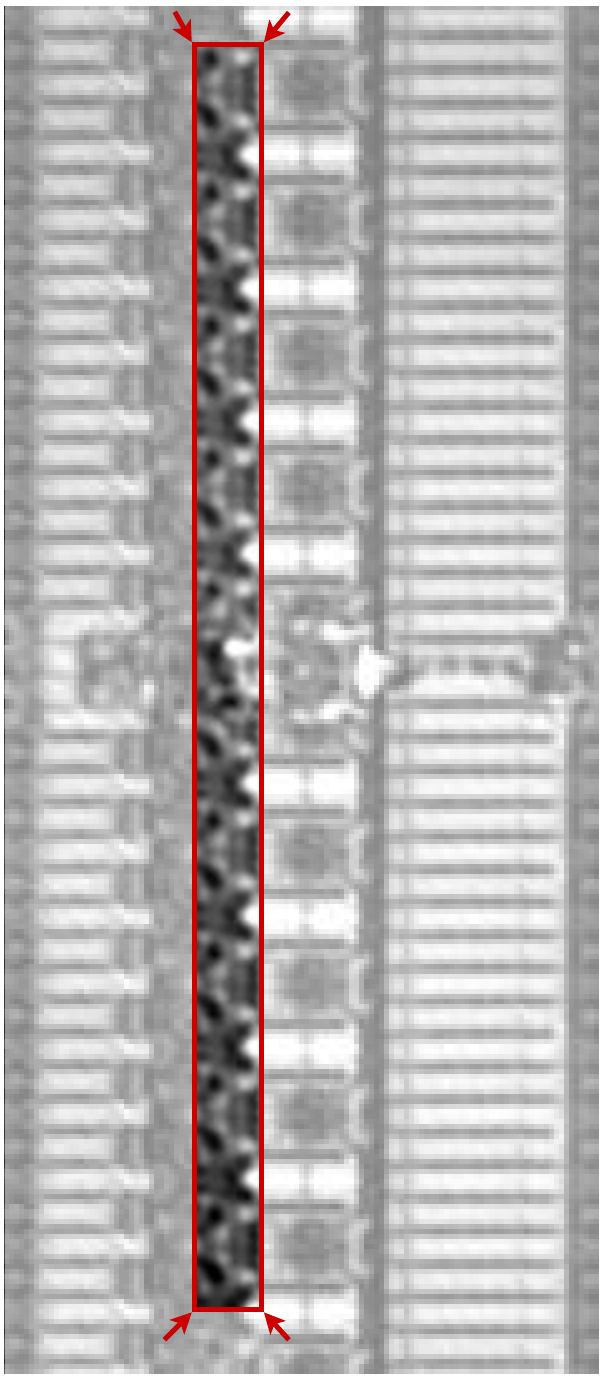}}
	
	\subfloat[LLSI image]{\includegraphics[angle=90,width=0.45\textwidth]{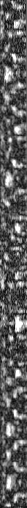}}
	
	\subfloat[Difference of two LLSI images]{\includegraphics[angle=90,width=0.45\textwidth]{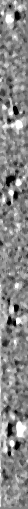}}
	\vspace{-1mm}
	\caption{Images of one Intel Cyclone IV LAB containing 16 registers.\label{fig:Cyclone-IV-target}}
\end{figure}

\subsubsection{Xilinx Kintex Ultrascale BBRAM}

As first and simple target we chose the battery-backed RAM (BBRAM) of a Xilinx Kintex Ultrascale FPGA, which is used for storing a 256-bit bitstream decrpytion key.
In principle, BBRAM is identical to common SRAM -- except that it is designed to be powered via battery over a long period.
Therefore, BBRAM cells are susceptible to optical SCA attacks.
In the literature it has been shown that the key from this device family can be extracted using TLS~\cite{lohrke_key_2018}.

The FPGA, which is manufactured in a 20\,\si{\nano\meter} technology, is mounted on an AVNET development board (AES-KU040-DB-G).
The flip-chip package of the FPGA allows direct access to the silicon backside of the chip.
For conducting TLS measurements, the current preamplifier is connected to the battery rails of the chip and the main power supply is switched off.
The bias voltage of the amplifier supplies the BBRAM during the TLS measurement.
For programming a new key, the FPGA has to be powered by the development board's power supplies.
To fully automate the programming and measurement process, we made use of the supplies' \mbox{PMBus} interface, allowing to switch the power on and off programmatically via a microcontroller (using the TI PMBus library~\cite{texasinst_msppmbus_2015}).
Consequently, for programming a new key, the power supplies are switched on, a key is programmed via JTAG and the Xilinx Vivado TCL interface~\cite{xilinxinc_vivado_2019}, and the power supplies are switched off again.
Note that during the whole process, the BBRAM voltage is supplied by the current preamplifier.
Fig.~\ref{fig:BBRAM-target-images} showcases images of the BBRAM area captured with our setup.
Although the chip is manufactured in a 20\,\si{\nano\meter} technology, the size of one memory cell is around 2.8\,\si{\micro\meter}\,$\times$\,3.1\,\si{\micro\meter}, which can be explained by leakage current considerations~\cite{lohrke_key_2018}.

\subsubsection{Texas Instruments MSP430 SRAM}

As second and more flexible target, we chose the freely programmable 1024-byte SRAM of a Texas Instruments MSP430 microcontroller.
The chip is manufactured in a 180\,\si{\nano\meter} technology with an SRAM cell size of approximately 2.5\,\si{\micro\meter}\,$\times$\,1.9\,\si{\micro\meter}~\cite{kiyan_comparative_2018}.
The literature shows that the SRAM content of this device can be extracted using TLS and LLSI~\cite{kiyan_comparative_2018}.
For our experiments, we chose to conduct LLSI measurements, as 
TLS is only possible while the device is in a low-power mode, which is not the case for LLSI.
Hence, LLSI can be considered a more powerful technique in this case.

To access the chip backside, the device had to be opened and soldered backside-up on a custom PCB.
Note that polishing or thinning the silicon backside was not necessary%
.
For modulating the power supply of the SRAM memory, we made use of the VCORE pin, which provides access to the internally generated core voltage of the microcontroller.
To this pin, we connected our modulator circuit, consisting of a voltage regulator whose feedback path is modulated using a laboratory frequency generator with a sinusoidal wave.
For programming the SRAM content during the automated measurements, we used an Olimex JTAG debugger (MSP430-JTAG-TINY-V2), controlled by a Python script using the MSPDebug command line tool~\cite{beer_dlbeer_2020}.
During the whole LLSI measurement, the debugger is left connected and switched on.
Fig.~\ref{fig:MSP430-target-images} showcases images of the SRAM area captured using our setup.

\subsubsection{Intel Cyclone IV Registers}

As the third target, we chose the registers of a Intel Cyclone IV FPGA.
The FPGA consists of 392 identical logic array blocks (LABs), each comprised of 16 logic elements (LEs), whereas every LE contains one register cell.
The chip is manufactured in a 60\,\si{\nano\meter} technology.
We had to open the package and solder the chip backside-up on a custom PCB for accessing the chip's backside.
To modulate the supply voltage for conducting LLSI, we used a voltage regulator (TI TPS7A7001) and modulated its feedback path with a sinusoidal wave.
We created a logic design that updates the register values when applying an external clock with precomputed randomly chosen values during the automated measurements.

By subtracting two LLSI images with different data, we found the LAB's area containing the registers.
To reduce the measurement time, we covered only that area with the automated measurements.
Consequently, one response image contains one logic array block, and therefore 16 registers, see Fig.~\ref{fig:Cyclone-IV-target}.
From the difference images, we could also estimate the memory cell size to around 7\,\si{\micro\meter}\,$\times$\,9\,\si{\micro\meter}.

\begin{table*}[htb]
\centering
{\renewcommand{\arraystretch}{1.25}%
{\footnotesize
\begin{tabular}{|c|>{\centering}p{.95cm}|>{\centering}p{0.85cm}|c|>{\centering}p{2.2cm}|>{\centering}p{1.8cm}|c|>{\centering}p{1.4cm}|>{\centering}p{1.25cm}|}
	\hline 
	\textbf{Target} & \textbf{\# Mem. bits} & \textbf{\# Key bits} & \textbf{Technique} & \textbf{Image dimensions} & \textbf{Lens and scanner zoom} & \textbf{\# Images} & \textbf{Time/Image (mm:ss)} & \textbf{Total time (hh:mm)}\tabularnewline
	\hline 
	\hline 
	{BBRAM} & {288} & {256} & {TLS} & {985\,px\,$\times$\,407\,px} & {50$\times$\,($\times$2)} & {578} & {02:02} & {19:35}\tabularnewline
	\hline 
	{MSP430 (zeroized)} & {8192} & {512} & {LLSI} & {503\,px\,$\times$\,355\,px} & {50$\times$} & {433} & {13:00} & {93:49}\tabularnewline
	\hline 
	{MSP430 (randomized)} & {8192} & {512} & {LLSI} & {503\,px\,$\times$\,355\,px} & {50$\times$} & {821} & {13:00} & {177:53}\tabularnewline
	\hline 
	{FPGA Registers} & {16} & {16} & {LLSI} & {509\,px\,$\times$\,28\,px} & {50$\times$\,($\times$2)} & {568} & {2:40} & {25:17}\tabularnewline
	\hline 
\end{tabular}
}
}
\caption{Overview of devices under test and the captured images in automated measurements.\label{tab:Overview-of-captured}}
\end{table*}

\section{Results\label{sec:Results}}

In this section we apply our deep learning based approach on the response images captured with the automated setup.
For all experiments, we first reduced the drift -- caused by mechanical instabilities of our setup -- between the images in the dataset.
For this, we calculated the offset between the optical image captured along with each response image and one fixed optical image by means of an elastic transformation using the MATLAB image processing toolbox.
Then we applied the transformation to the corresponding response image.
For the sake of simplicity, we will in the following refer to the response images only as ``images''.
To encourage others working with our data, we made all images captured in the context of this work available online.\footnote{%
\ifthenelse{\boolean{cameraready}}{
\url{http://dx.doi.org/10.14279/depositonce-11354}%
}
{
In the final publication, a DOI will be provided here.%
}
}

\subsection{Key Extraction from BBRAM\label{subsec:BBRAM}}

Using the automated setup, we have captured over 500 TLS images of the BBRAM containing randomly chosen keys, see Tab.~\ref{tab:Overview-of-captured} for details.
The memory cells' locations within the image become visible when subtracting two TLS images containing different keys, see Fig.~\ref{subfig:BBRAM-target-images_diff}.
The relatively large spots indicate that the memory cells cover many pixels, and therefore, we downsized the images with a factor of 0.4 before using them for training.
We first investigated if it is possible to extract single key bits from the images (Section~\ref{subsec:BBRAM_single_bits}).
Further, we examined ways for reducing the required time for learning (Section~\ref{subsec:BBRAM_parallel}) and the number of images in the training dataset (Section~\ref{subsec:BBRAM_num_imgs}).
Finally, we constructed an optimized attack approach from our findings (Section~\ref{subsec:BBRAM_attack_approach}).

\subsubsection{Learning single bits\label{subsec:BBRAM_single_bits}}

For the first experiments, we fed images containing the entire BBRAM area into the network (cf. Fig.~\ref{fig:BBRAM-target-images}).
For the CNN, we used a simple VGG-like structure, consisting of just two convolutional layers, followed by a pooling layer, and a FC network with one hidden layer (512 neurons), a dropout layer (rate 0.2), and an output layer with one neuron.
For the model summary, see Fig.~\ref{fig:model} in the Appendix.
For all experiments in this work, we used the Adam optimizer with an initial learning rate of $0.001$, binary cross-entropy loss functions, and rectifier activation functions. 
We randomly split the available images into training (70\%), validation (15\%), and test (15\%) datasets.
Further, we applied a batch size of $8$ images and set the number of steps per epoch to the number of images in the training dataset divided by the batch size.
To deal with the relatively small datasets, we augmented the images by means of an affine transformation with a random rotation of 2 degrees, a width/height shift of 1 pixel, and a shear of 2 degrees.

\begin{figure}[tb]
	\centering{}
	\setlength{\axwidth}{\linewidth}
	\setlength{\axheight}{3.5cm}
	\input{figs/results/ultrascale/single_bit_acc.tex}
	\vspace{-6mm}
	\caption{Training and validation accuracy when learning a single bit of the BBRAM key from the full image.
	\label{fig:BBRAM_single}}
\end{figure}
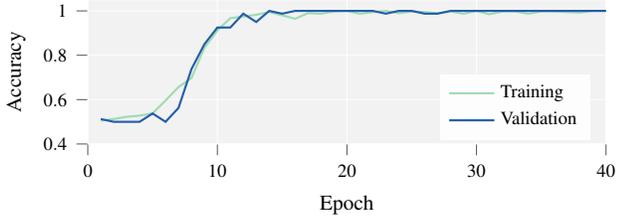

The results show that the network can quickly learn one bit of the key, see Fig.~\ref{fig:BBRAM_single}. 
We repeated the experiment for 50 randomly chosen bit positions of the key, and recognized, that not all networks lead to a test accuracy of 100\%.
Therefore, we repeated the network training five times per key bit for different splits of the dataset.
In most runs (at least 3 out of 5), we achieved a test accuracy of 100\%.
The reasons for some networks to perform better and some worse could be the relatively small number of training images and the random initialization of the networks' weights.
To make predictions of the secret more reliable, an ensemble learning strategy can potentially be used, for instance, by considering the models from multiple runs in a majority voting fashion.
Training a network for one bit took around 180 seconds per run, which -- depending on the number of runs -- can lead to a training time of some hours to a few days.

\subsubsection{Learning bits in parallel\label{subsec:BBRAM_parallel}}

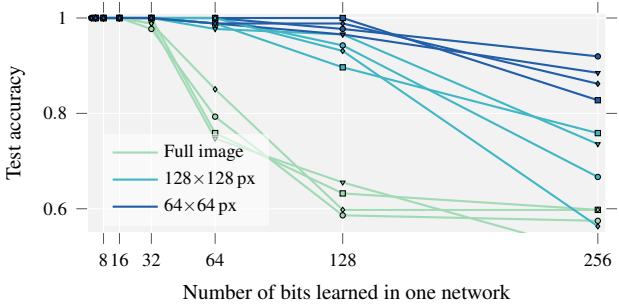
\begin{figure}[tb]
	\centering{}
	\setlength{\axwidth}{\linewidth}
	\setlength{\axheight}{4.5cm}
	\input{figs/results/ultrascale/bits_in_parallel_test_acc_4_bits_up_to_256.tex}
	\vspace{-1mm}
	\caption{Test accuracies for four bits of the BBRAM key when trying to learn multiple bits in parallel with one network. Shown values depict the maximum out of 3 runs.\label{fig:BBRAM_parallel_allImgs} }
\end{figure}

\begin{figure}[tb]
	\centering{}
	\setlength{\axwidth}{\linewidth}
	\setlength{\axheight}{4.5cm}
	\input{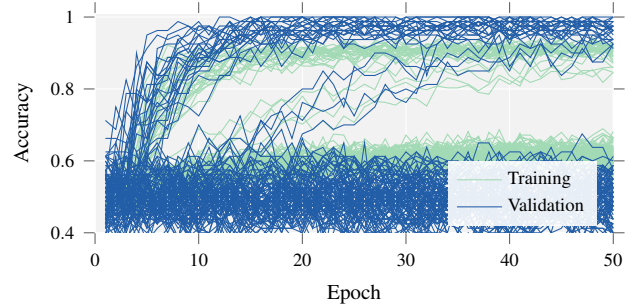}
	\vspace{-6mm}
	\caption{Training history when learning 128 bits of the BBRAM key on a \sixyfourtimespx section. The bits contained in the section converge to 100\% accuracy, and therefore, can be clearly separated from the others.\label{fig:BBRAM_parallel_allImgs_hist}}
\end{figure}

To speed up model training for all key bits, we added more output neurons to the network to learn multiple key bits in parallel.
For this, we randomly chose bit positions from the key and checked if we can achieve a simultaneous test accuracy of 100\% for all bits.
This was the case for up to 4 key bits per network, when training for the same number of epochs as before on the full image, which leads to a 4$\times$ speedup in training time. Above 4 bits, the test accuracy was decreasing significantly.
Increasing the number of convolutional and FC layers did not improve the prediction accuracy.
Further, we noticed that the achieved performance depends on the spatial distance between the memory cells learned in parallel.
When trying to learn cells in close vicinity, the per-bit accuracy is higher than with randomly chosen memory cells.

To further increase the number of bits learned in parallel, we reduced the network's data input dimensions by breaking the images into sections, and training one network for each section.
Now not all key bits are contained within one section, and consequently, an accuracy of around 50\% might indicate that the section does not contain the corresponding bit.
Therefore, we picked four bits that are contained in a specific \hundredtwentyeighttimespx and \sixyfourtimespx section, and tried to learn up to 256 bits of the key in parallel from differently sized sections.
The results confirm that a smaller section size leads to a higher accuracy.
We could achieve a test accuracy of 100\% for all four bits contained in the section when trying to learn up to 32 bits in parallel, see Fig.~\ref{fig:BBRAM_parallel_allImgs}.
Although not reaching a very high test accuracy, the network for learning 128 bits in parallel can clearly separate bits that are contained in the section from bits that are not, see Fig.~\ref{fig:BBRAM_parallel_allImgs_hist}.
A few bits achieve a higher validation accuracy only in later epochs, presumably because they are not fully contained in the image section, and therefore, are harder to learn.

To sum up, this experiment has shown two things.
Firstly, breaking the images into smaller sections can increase the achieved accuracy of the model.
Secondly, the bits' rough locations on the image can be found very efficiently, by learning many bit positions of the key in parallel.

\subsubsection{Reducing the number of required images\label{subsec:BBRAM_num_imgs}}

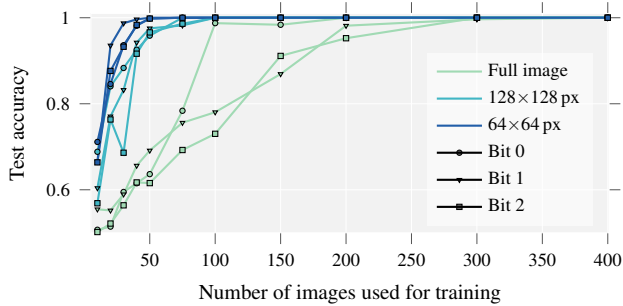
\begin{figure}[tb]
	\centering{}
	\setlength{\axwidth}{\linewidth}
	\setlength{\axheight}{4.5cm}
	\input{figs/results/ultrascale/single_bits_num_imgs.tex}
	\vspace{-6mm}
	\caption{Learning one bit of the BBRAM key per network from differently sized sections, with respect to the number of images used for training. The experiment was repeated for three key bits.\label{fig:BBRAM_num_imgs}}
\end{figure}

We expect the cost of using the FA microscope, i.e., for capturing the images, to be in orders of magnitude higher than the cost for training the CNNs.
Therefore, we consider the required number of training images as the limiting factor regarding the attack costs.
Consequently, we tried reduce the number of samples used for training to a minimum, while still being able to extract the secret.
For this, we again tried to learn only single bits per network, and repeated the experiment for three bits of the key on different image section sizes.
In a nutshell, the results indicate that training on a smaller section size requires a smaller test dataset, see Fig.~\ref{fig:BBRAM_num_imgs}.
Remarkably, to learn a single bit from a \sixyfourtimespx section with 100\% accuracy demands only 50 training images.

\subsubsection{Optimized attack approach\label{subsec:BBRAM_attack_approach}}

From the above findings, we can now develop an attack approach that is adaptable to constraints like the amount and quality of available images.
We propose a two-step divide-and-conquer approach as follows.
First, for finding the bits' coarse locations, networks are trained on many bits in parallel for small sections of the original image.
Note that high test accuracies are dispensable in this case, since it is only of interest whether a bit is learnable or not.
Once the coarse location of each bit is found, networks for each bit (or small groups of bits) can be trained on the corresponding sections.

\begin{figure*}[tb]
	\centering
	\includegraphics[width=0.65\textwidth]{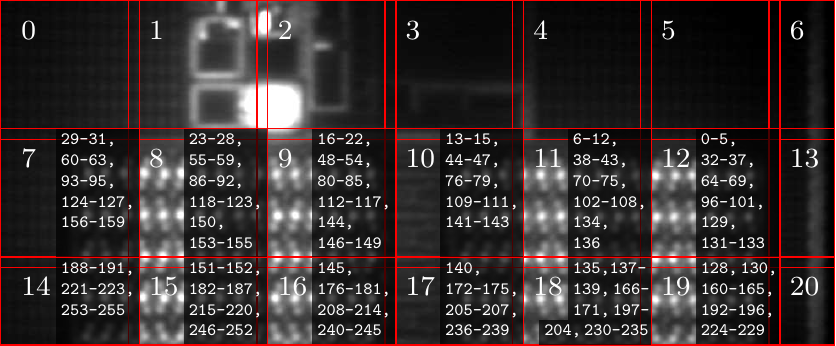}
	\caption{BBRAM  memory area split into \sixyfourtimespx sections.
	The small numbers indicate the localized key bits for each section (most significant bit = 0).\label{fig:BBRAM_attack_fractions}}
\end{figure*}

\begin{figure}[htb]
	\centering{}
	\setlength{\axwidth}{\linewidth}
	\setlength{\axheight}{4cm}
	\input{figs/results/ultrascale/bits_parallel_few_imgs.tex}
	\vspace{-2mm}
	\caption{Trying to learn the first 128 bits of the BBRAM key in parallel on section 12 (see Fig.~\ref{fig:BBRAM_attack_fractions}) with only 150 images used%
	.\label{fig:BBRAM_attack_localization}}
\end{figure}
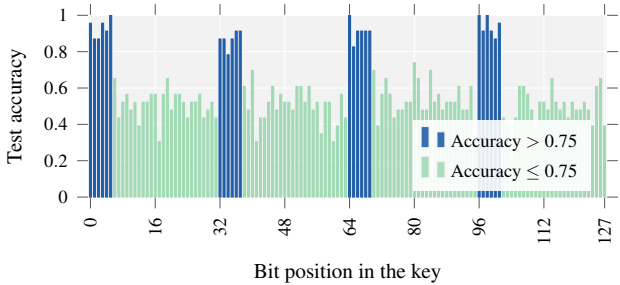

\vspace{1mm}
\noindent
\textbf{Localization\quad}
We chose to reduce the training dataset to only 150 images to better reflect a real attack scenario in which capturing time is expensive, resulting in a dataset acquisition time of 5~hours.
We then trained networks for 128 bits in parallel on \sixyfourtimespx sections of the images with 5\,\si{\pixel} overlap at every side, resulting in 21 sections, see Fig.~\ref{fig:BBRAM_attack_fractions}.
We ran every training three times and selected the most promising section for each bit by first filtering for test accuracies above 75\% and then picking the section with the highest number of successful runs.
For instance, some of the key bits between 0 and 127 could be learned in section 12, see Fig.~\ref{fig:BBRAM_attack_localization}.
The algorithm found bit numbers 0-5, 32-37, 64-69, 96-101, 129, and 131-133\footnote{Numbering with most significant bit first.}.
This matches with the memory mapping already discovered in \cite{lohrke_key_2018}.
Note that the bits 128-133 seem to reside directly in the overlap region of sections 12 and 19, and therefore, some bits could be better learned in section 12, and some in 19.
We could successfully find the corresponding section for every bit of the key.
One training run took 133~seconds, which results in a total localization time of 4:42~hours.

Additionally, we reduced the dataset to 100 images and trained networks only for 64 bits in parallel on \sixyfourtimespx sections.
The experiment delivered the same localization results as before, with a slightly shorter training time (4:24 hours).
Consequently, we believe that tweaks and optimizations can reduce the number of required images even further. 

\vspace{1mm}
\noindent
\textbf{Prediction\quad}
Once all bits' rough locations are known, at most one network training per key bit is necessary to predict all bits with high accuracy.
The previous results indicate that there is a trade-off between training time and training dataset size.
Training one network on a \sixyfourtimespx image section for one key bit with a dataset consisting of 100 images takes around 30 seconds, resulting in a total training time for all
bits of the key of 2:08 hours (for one run per model).
To increase the bit prediction accuracy, multiple training runs can potentially be combined in an ensemble learning strategy with only a linear increase in training time.

\subsection{Key Extraction from Microcontroller SRAM\label{subsec:msp430}}

On the microcontroller SRAM as our most flexible target, we evaluated two scenarios.
In scenario 1 (Section~\ref{subsec:Simple-(rest-zeroized)}), we programmed a randomly chosen key into 512\,bits of the 1\,kB (= 8192\,bits) memory at the addresses \texttt{0x10}\,--\texttt{\,0x4f}, while keeping the rest of the memory zeroized.
This scenario corresponds to the BBRAM target, except for the smaller memory cell sizes and the more distributed memory cells holding the key.
In the scenario 2 (Section~\ref{subsec:Advanced-(rest-randomized)}), the entire memory content is randomized.
Again, we consider the same 512 bits of the memory to be the key which should be extracted. This scenario simulates a high amount of irrelevant information in the measurement, caused by other activities on the chip or intended obfuscation.

\subsubsection{Scenario 1: Rest zeroized\label{subsec:Simple-(rest-zeroized)}}

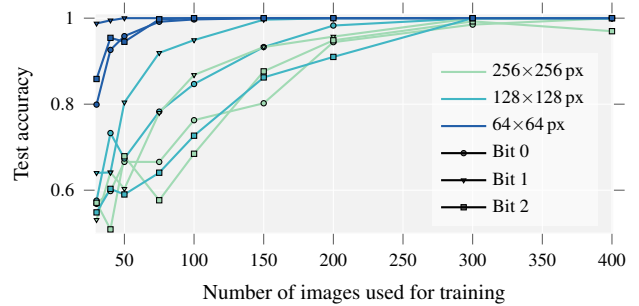
\begin{figure}[tb]
	\centering
	\setlength{\axwidth}{\linewidth}
	\setlength{\axheight}{4.5cm}
	\input{figs/results/msp/single_bits_num_imgs.tex}
	\vspace{-6mm}
	\caption{SRAM scenario 1 -- Learning one key bit per network from differently sized sections with respect to the number of images used for training. The experiment was repeated for three bit positions.\label{fig:MSP_num_imgs_simple}}
\end{figure}

We captured over 400 images for this scenario, see Tab.~\ref{tab:Overview-of-captured} for details.
Fig.~\ref{subfig:MSP430-target-images_diff} indicates that the images are not as clear as the BBRAM images.
The reason is that we did not use an extra 2$\times$ scanner zoom like for the BBRAM, because we wanted to fit the whole memory into one image.
Furthermore, the memory cells are slightly smaller than those of the BBRAM.
The difference image of two different keys (see Fig.~\ref{subfig:MSP430-target-images_diff}) indicates that the key is distributed over large parts of the memory, and therefore, nearly the whole image must be considered for extracting the key bits.
We first investigated the required number of images for learning one bit, see Fig.~\ref{fig:MSP_num_imgs_simple}.
The results show that around 100 images are sufficient to reach 100\% test accuracy for a \sixyfourtimespx section.
For a \hundredtwentyeighttimespx section, already around 400 images are required to achieve a test accuracy of 100\%.
The network architecture and setup working best is identical to the setup used for the BBRAM key extraction (Section \ref{subsec:BBRAM_single_bits}).

For the localization step, we split the images into \sixyfourtimespx sections, resulting in 54 sections, see Fig.~\ref{fig:msp_sections_localization} in the Appendix.
For every section, we trained models on 128 key bits in parallel. %
We could localize all 512 key bits by using 300 images from the dataset.
The results are shown in Tab.~\ref{tab:msp_localization} in the Appendix.
Note that the number of images can be reduced when accepting longer localization times -- by learning less bits in paralel.

\subsubsection{Scenario 2: Rest randomized\label{subsec:Advanced-(rest-randomized)}}

\begin{figure}[tb]
	\centering
	\includegraphics[height=.8\linewidth, angle=90]{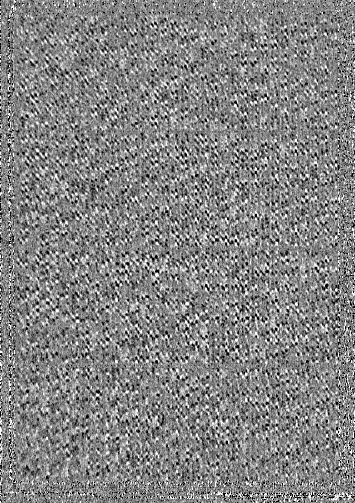}
	\caption{SRAM scenario 2 -- Difference between two LLSI images with the entire memory randomized (image rotated clockwise by 90°).\label{subfig:MSP430-target-images_diff-rand}}
\end{figure}

In the previous scenario, there was not much noise present in the images.
However, on a real target, surrounding memory cells might not always hold the same value. 
Therefore, we randomized the entire memory content for scenario~2.
The subtraction of two LLSI images shows that no longer any area of interest can be recognized, see Fig.~\ref{subfig:MSP430-target-images_diff-rand}.
We assumed that this scenario is harder to learn, and therefore, captured over 800 images, see Tab.~\ref{tab:Overview-of-captured} for details.

\begin{figure}[tb]
	\centering
	\setlength{\axwidth}{\linewidth}
	\setlength{\axheight}{4.5cm}
	\input{figs/results/msp/single_bits_num_imgs_randomized.tex}
	\vspace{-6mm}
	\caption{SRAM scenario 2 -- Learning one bit per network from differently sized sections while the whole memory content is randomized. Experiment is repeated for three bit positions.\label{fig:MSP_num_imgs_advanced}}
\end{figure}
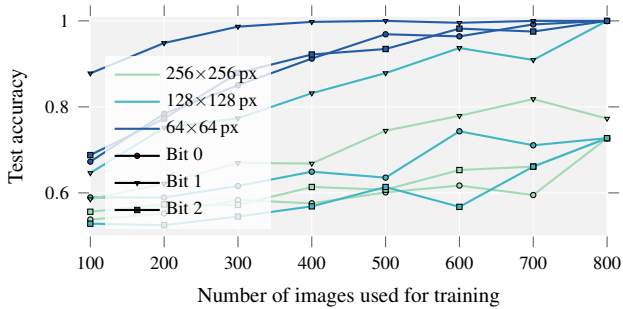

As before, we first investigated how many images are required to extract single bits.
The results show that -- compared to scenario 1 -- eight times more images are necessary to achieve a test accuracy of 100\%, see Fig.~\ref{fig:MSP_num_imgs_advanced}.
Interestingly, only one of the three bits achieves a test accuracy of 100\% for \hundredtwentyeighttimespx sections (Bit 1). Also for \sixyfourtimespx sections, the other two bit positions (Bit 0 and Bit 2) show clearly worse accuracies.
For the other bits and larger sections, the number of images seems to be insufficient to achieve a very high test accuracy.

When using 400 images for the localization and learning 128 bits per network, we could map 91\% of the bits to the same sections as in scenario 1.
In other words, 45 out of 512 bit positions were not found in their correct section. Therefore, we ran the same experiment using 800 images.
Although still 12 bit positions were not mapped to the same sections as in scenario 1, they could be located in a neighboring section.
The reason is that those bits seem to be located in the overlap region of the two sections.
The results show that a high level of irrelevant information increases the amount of required images significantly.
Nevertheless, extracting the key is still possible when spending enough time on measurements. %

\subsection{FPGA Register Content Extraction}

Our dataset for this target consists of more than 500 images, each containing one logic array block (LAB) with 16 register bits, see Tab.~\ref{tab:Overview-of-captured} for details.
Note that not all images show the physically same registers on the chip, but instead instances of the same logic layout.
Therefore, if the bit values can be learned in our experiment, the resulting predictor can be used to extract data from all LABs distributed over the FPGA.

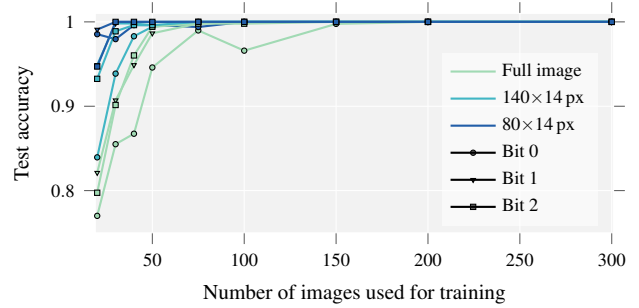
\begin{figure}[tb]
	\centering
	\setlength{\axwidth}{\linewidth}
	\setlength{\axheight}{4.5cm}
	\input{figs/results/cyclone/single_bits_num_imgs.tex}
	\vspace{-6mm}
	\caption{FPGA registers -- Learning one bit per network from differently sized sections with respect to the number of images used for training and validation. The experiment was repeated for three bit positions.\label{fig:CYC_num_imgs}}
\end{figure}

\begin{figure*}[tb]
	\centering
	\includegraphics[width=.8\textwidth]{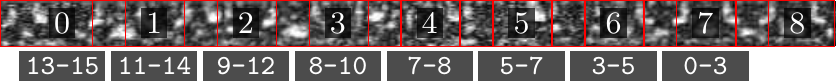}
	\caption{Sections of the FPGA register area for localizing the bits' rough positions. The number ranges indicate the bit positions of the secret localized in the respective section.\label{fig:cyclone_sections_localization}}
\end{figure*}

Like for the other targets, we investigated the influence of the training dataset size on the test accuracy when training networks on a single bit of the secret.
The results indicate that -- depending on the section size of the images -- at most 150 images are required to achieve a test accuracy of 100\%, see Fig.~\ref{fig:CYC_num_imgs}.
Although the bits can already be learned from the full images with a low number of training samples, we further split the images into smaller sections to localize the individual bits in more detail.
Fig.~\ref{fig:cyclone_sections_localization} shows the results for splitting the images into 8 sections, which already gives very precise information on the bits' position.

\section{Discussion\label{sec:Discussion}}

\subsection{Scalability of Data Extraction\label{subsec:scalability}}
One important aspect is the scalability of our approach towards the extraction of larger chunks of data and implementations employing classical countermeasures against SCA attacks (e.g., Boolean masking).

In our experiments on the MSP430 microcontroller (Section \ref{subsec:msp430}), we have captured images of the full 1024-byte SRAM with randomly chosen content.
We have defined 64 bytes in a fixed address range as the key bits, and have shown that all bits can be localized and extracted from the images.
Since we could have chosen any other address range within the memory as key storage, it will also be possible to extract the entire memory content from the images with only a linear increase in extraction time.
Consequently, we expect our approach to work also on larger chunks of data with only a linearly growing effort.

One might ask if the approach is also applicable when the key is not present in plaintext on the chip.
Examples for implementations that do not require a key in plaintext are masked versions of cryptographic cores %
that work on shared forms of the key~\cite{ishai_private_2003}.
Previous work has already shown that all key shares can be extracted using laser-assisted SCA when all potential memory/register locations are known to the adversary: either by direct readout or with the help of a SAT solver~\cite{krachenfel_realworld_2020}.
In this work, we assume zero knowledge about the memory locations on the chip.

In preliminary experiments, we presume a 2-share Boolean masking of the key, meaning that the unmasked key can only be obtained by XOR'ing two values stored in the memory.
We artificially created the masking on the available dataset by defining pairs of memory locations as the shares.
In other words, on a memory snapshot containing $N$ bit values $b_0\:...\:b_{N-1}$, one key bit $k$ for a 2-share masking is $k=b_x \oplus b_y$ $(0 \leq x,y < N, x \neq y)$.
During the profiling phase, only the unmasked key $k$ is known to the adversary.
We trained models on a 128\,\si{\pixel}\,$\times$\,128\,\si{\pixel} section of the BBRAM images containing $N=47$ bits, and achieved 100\% test accuracy for all exemplarily tested bit combinations (e.g., for $(x,y) \in \{(0,4),(1,8),(32,66)\}$, cf. Fig.~\ref{fig:BBRAM_attack_fractions}).

Hence, the network has not only learned the memory locations of the individual shares, but also that the values have to be XOR'ed to obtain the unmasked key.
We used the same neural network structure as in all the other experiments presented in this work and observed that the model needs to be trained for more epochs than for the direct key extraction.
On the MSP430 microcontroller SRAM, we only had success on some bit combinations, and therefore, we believe that the network architecture will have to be adapted to work more reliably.
A more thorough exploration of masked data extraction can be conducted in the future using the data collected in this work.
In summary, also the unmasked key of a masked implementation can be extracted using our laser-assisted SCA approach.

\subsection{Optical Resolution and Cell Size}

Optical resolution is defined as the ability of an optical system to differentiate between two closely spaced objects.
Because of constant decrease in feature sizes -- now reaching down to the 5\,\si{\nano\meter} node, optical resolution has been a growing concern for the FA community.
Debugging the root cause of a failure can require to resolve adjacent minimum size transistors from each other, which might be challenging when we think of the most recent technology nodes.
Tools such as the solid immersion lens (SIL) and visible light source systems~\cite{beutler_visible_2015,boit_contactless_2015} have been introduced to overcome this problem.
It has been shown that a SIL can improve the optical resolution down to approximately 200\,\si{\nano\meter}, enabling optical probing even for 10\,\si{\nano\meter} technology nodes~\cite{boit_ic_2016,vonhaartm_optical_2015}.
With our setup, we can achieve a laser spot diameter of approximately 1\,\si{\micro\meter} without a SIL.
Laser power at the center of the spot is the strongest and decreases exponentially through the edge.

The transistors in the SRAM cells are often designed to be larger than those used in the logic part of the chip to avoid off-leakage current related data loss.
Although the DUTs in our experiments were manufactured in technology nodes down to 20\,\si{\nano\meter}, the contained memory cells were larger than expected.
Among the DUTs that we have used, the smallest cell size is 2.5\,\si{\micro\meter}\,$\times$\,1.9\,\si{\micro\meter} in MSP430 which is still larger than the 1\,\si{\micro\meter} laser diameter.
In the case of the Xilinx Ultrascale BBRAM, the cell size is even larger, although the technology size is much smaller.
This shows that cell sizes do not always proportionally scale with the technology nodes, but cell size scaling also depends on many other parameters such as current leakage or supply voltage.
The designers have to keep the transistor sizes bigger to maintain the circuit performance and the yield.
In addition to that, while logic density continues to double in every technology generation, the memory cell size shrink cannot keep up the trend at the same pace.
As a result, the memory density increase remains less than double at every new technology node~\cite{maheshwari_memory_2014}.
The limiting factor appears to be lithography and the cost associated with it~\cite{keshavarzi_directions_2014}. 

The question whether it is possible to extract logic states from memory cells that are smaller than the laser spot size can not be answered trivially.
While for FA purposes it might be important to target only a single transistor, for our approach it is only important that the response image differs in some way between the logic states \texttt{0} and \texttt{1}.
As a matter of fact, the distances and the positions of the opposite state transistors with respect to each other are more important than the transistor sizes.
For our DUTs, we do not know the exact memory structure, and we have not tested our approach on memories other than presented in this work.  However, this is among our future research interests.

For the optical SCA techniques used in this work, the laser beam is scanned over the device pixel-wise.
When reducing the pixel size to values smaller than the laser spot diameter, for every pixel the superimposed signal/response originating from multiple transistors or memory cells will be captured. Consequently, the resulting response image will be noisy.
We suppose that image processing tools like CNNs can be used to recover the logic state from the interfering signals.
To the best of our knowledge, this has not been investigated in the hardware security community, and therefore, it is among our planned future works.
In conclusion, the optical resolution might be a challenge when going to memories with smaller cell sizes and higher cell densities.
However, we assume that optical contactless probing will continue to be present for a while due to the reasons mentioned above.

\subsection{Chip Access\label{subsec:discussion_chip_access}}

All the above mentioned SCA techniques are performed through the chip backside, which means that the attacker should have access to the bulk silicon.
Since many modern ICs are manufactured in flip-chip packages, optical attacks are easy to conduct and often even do not require extra preparation steps.
For instance, the Xilinx Kintex Ultrascale FPGA is shipped in a bare-die flip-chip package, and therefore, no preparation was needed for silicon access.
In contrast, the packages of the other targeted devices had to be opened and soldered back-side up on a custom PCB for accessing the backside, which makes it a semi-invasive attack.
Nevertheless, it should be noted that for technology nodes of 20\,\si{\nano\meter} and below, flip-chip packages are becoming more prevalent due to performance, size and cost issues~\cite{tong_advanced_2013}.

\subsection{Attack Cost and Time Expenditure}

Our investigations have shown that -- depending on the area of interest on the chip and the imaging resolution -- several hours to a few days have to be spent for automated measurements on the training device.
This time is presumably the most costly period when conducting the proposed attack. This is not a challenge when the attacker owns a setup for conducting the attacks.
The tools for conducting TLS and LLSI cost around \$1M, whereas a setup for conducting only TLS can be acquired for around \$100k~\cite{krachenfel_evaluation_2020}.
Since a laser scanning microscope is common equipment in FA labs around the globe, a suitable setup can also be rented for about 300\$/h including an operator.
Consequently, we can calculate the costs for acquiring the images as given in Tab.~\ref{tab:Overview-of-captured} to \$509 for the BBRAM (50 images), around \$6.5k for scenario 1 (100 images) and \$52k for scenario 2 on the microcontroller SRAM, and \$667 for the FPGA registers (50 images).
Note that due to the mostly automated measurements, which can also run unsupervised during the night, those fares could presumably be reduced.
Furthermore, a more stable optical setup would avoid the need for a frequent auto-focus and drift correction, and therefore, can potentially reduce the measurement times according to our estimations by up to 50\%.
Although we agree that the costs are still high for some scenarios, we would like to stress that the gathered model is applicable to all devices of a device series, and can extract the secrets contained in multiple devices.

\subsection{Key Control\label{subsec:discussion_key_control}}

One might argue that it is not always true that the adversary can program different keys into the NVM on a training device, for instance, when one-time programmable (OTP) memories like e-fuses or ROMs are used.
We admit that such keys cannot be extracted using our approach.
However, in many applications a OTP memory only stores a key-decryption-key, which is used to decrypt other application keys contained in reprogrammable NVMs.
This makes the system more flexible and keys can be updated together with the device’s firmware.
Since the application keys will be decrypted by some cryptographic core on the device, they will in the end also be stored in registers on the chip.
We have shown that this kind of application keys can be targeted using our approach.

\subsection{Potential Countermeasures\label{subsec:discussion_countermeasures}}

When looking for potential countermeasures, one should keep in mind that potentially many different FA techniques can be used to read out the logic states of the device under attack.
Therefore, a countermeasure should at best protects against all possible attack techniques.
In other words, there exist various countermeasures that are effective against some FA-based attack techniques, but do not necessarily prevent other methods.

One technique proposed for protecting semiconductor intellectual property is IC camouflaging~\cite{shakya_covert_2019,rajendran_security_2013}.
Therefore, one might ask if camouflaging also can protect against memory readout.
The idea behind camouflaging is to insert logic gates whose functionality cannot be extracted by delayering the chip and applying imaging techniques like SEM.
However, since optical techniques rely on interactions with the actual transistors, they can still recognize the function of the camouflaged gates~\cite{shakya_covert_2019}.
In other words, it would be possible to extract the logic states using activity maps of the circuit.
Consequently, camouflaging does not seem to be an appropriate countermeasure.

The foremost requirement for our attack approach to succeed is access through the chip's backside.
Active backside coatings~\cite{amini_assessment_2018} can prevent the optical access to the chip's silicon by adding an opaque coating layer.
By actively checking the intactness of the coating, attempts to remove it can be detected.
Since removing the silicon substrate from the chip backside is necessary for conducting SEM- and FIB-based attacks, an active coating can also help in these cases.
However, to the best of our knowledge, there is no implementation of an active backside coating ready for mass production.

According to the preliminary results presented in Section \ref{subsec:scalability} on masking implementations, Boolean masking seems to increase the effort for the attacker, but does not prevent laser-assisted SCA to a sufficient degree.

\section{Conclusion\label{sec:Conclusion}}

Hardware attacks using sophisticated FA tools are often seen as too costly and time-consuming to pose a severe threat to modern ICs and SoCs.
Therefore, vendors usually rely on the complexity of the layout and tamper-proof memories to prevent key extraction.
However, for being used, every key will be cached into memory cells that are vulnerable to probing techniques, such as optical SCA.
In this work, we have shown that the automation of FA tools combined with deep learning techniques reduces the required effort by an adversary significantly.
We carried out highly automated measurements on three different hardware targets holding an attacker-controlled secret in their memories.
Besides, we have demonstrated how to fully extract the secret from the captured images without knowing the chip's layout, especially the memory cells' design, geometry, and exact location.
We believe that our approach has the potential to antiquate the expensive reverse engineering part of hardware attacks by offering a very targeted and generic procedure for key extraction, which can also be applied in the presence of largely irrelevant information and activities on the chip.
Hence, a great deal of attention has to be paid to this threat when designing new RoT devices for critical applications.
While, in this work, we presented an offensive application of our approach, it also can be utilized to assess the vulnerability of the products in the early stages of the design, and consequently, assist in finding the right defense techniques.

\ifthenelse{\boolean{cameraready}}{
\section*{Acknowledgment}
The work described in this paper has been supported in part by the Einstein Foundation in form of an Einstein professorship -- EP-2018-480, in part by the Deutsche Forschungsgemeinschaft (DFG -- German Research Foundation) under the priority programme SPP 2253 -- 422730034, and in part by the  German  Ministry  for  Education and Research as BIFOLD -- Berlin Institute for the Foundations of Learning and Data (ref. 01IS18025A).
The authors would also like to acknowledge Hamamatsu Photonics K.K. Japan and Germany for their help and support on the \mbox{PHEMOS }system.
}{}

\FloatBarrier

\begingroup
\setstretch{0.95}
\printbibliography
\endgroup

\appendix
\clearpage
\section*{Appendix}

\noindent
\begin{minipage}{\linewidth}
\makebox[\linewidth]{
	\includegraphics[width=.95\linewidth, trim=0 5.5cm 0 0, clip]{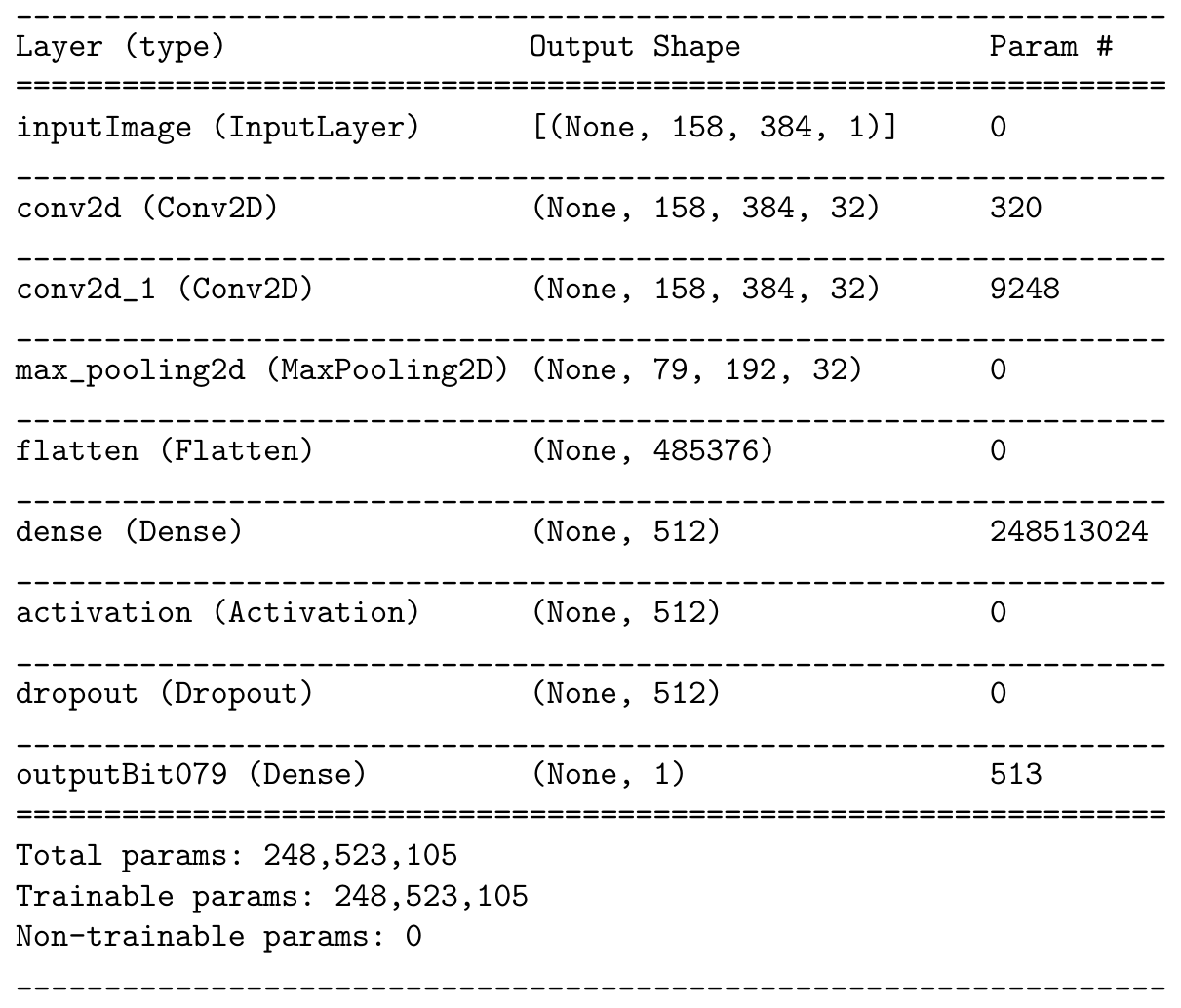}}
	\captionof{figure}{CNN model summary for the BBRAM experiments, here for learning bit 79 of the key.\label{fig:model}}
\end{minipage}
\vfill\null
\vspace{5mm}

\noindent
\begin{minipage}{\linewidth}
	\makebox[\linewidth]{
	\setlength{\axwidth}{.95\linewidth}
	\includegraphics[width=.95\linewidth]{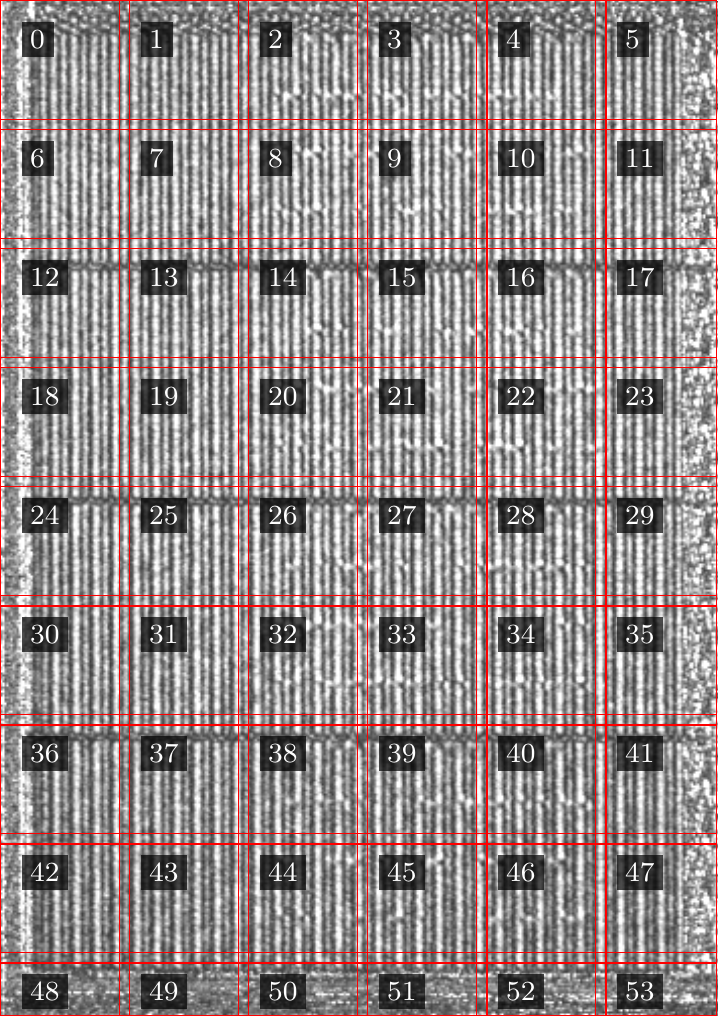}
	}
	\captionof{figure}{Sections of the MSP430's SRAM area used for localizing the bits.\label{fig:msp_sections_localization}}
\end{minipage}
\vfill\null

\noindent
\begin{minipage}{.49\textwidth}
	\makebox[\linewidth]{
{\renewcommand{\arraystretch}{1.25}%
{\footnotesize
\begin{tabular}{>{\centering}p{.85cm}|p{6.5cm}} 
\textbf{Section} & \textbf{Bit positions} (most significant bit first) \\ \hline
2 & 375, 383, 391, 399, 407, 415, 423, 431, 439, 447, 455, 463, 471, 479, 487, 495, 503, 511 \\ 
3 & 183, 191, 199, 207, 215, 223, 231, 239, 247, 255, 263, 271, 279, 287, 295, 303, 311, 319, 327, 335, 343, 351, 359, 367 \\ 
4 & 7, 15, 23, 31, 39, 47, 55, 63, 71, 79, 87, 95, 103, 111, 119, 127, 135, 143, 151, 159, 167, 175 \\ 
8 & 374, 382, 390, 398, 406, 414, 422, 430, 438, 446, 454, 462, 470, 478, 486, 494, 502, 510 \\ 
9 & 182, 198, 206, 214, 222, 230, 238, 246, 254, 262, 270, 278, 286, 294, 302, 310, 318, 326, 334, 342, 350, 358, 366 \\ 
10 & 6, 14, 22, 30, 38, 46, 54, 62, 70, 78, 86, 94, 102, 110, 118, 126, 134, 142, 150, 158, 166, 174, 190 \\ 
14 & 373, 381, 389, 397, 405, 413, 421, 429, 437, 445, 453, 461, 469, 477, 485, 493, 501, 509 \\ 
15 & 181, 189, 197, 205, 213, 221, 229, 237, 245, 253, 261, 269, 277, 285, 293, 301, 309, 317, 325, 333, 341, 349, 357, 365 \\ 
16 & 5, 13, 21, 29, 37, 45, 53, 61, 69, 77, 85, 93, 101, 109, 117, 125, 133, 141, 149, 157, 165, 173 \\ 
20 & 372, 380, 388, 396, 404, 412, 420, 428, 436, 444, 452, 460, 468, 476, 484, 492, 500, 508 \\ 
21 & 188, 196, 204, 212, 220, 228, 236, 244, 252, 260, 268, 276, 284, 292, 300, 308, 316, 324, 332, 340, 348, 356, 364 \\ 
22 & 4, 12, 20, 28, 36, 44, 52, 60, 68, 76, 84, 92, 100, 108, 116, 124, 132, 140, 148, 156, 164, 172, 180 \\ 
26 & 371, 387, 395, 403, 411, 419, 427, 435, 443, 451, 459, 467, 475, 483, 491, 499, 507 \\ 
27 & 179, 195, 203, 211, 219, 227, 235, 243, 251, 259, 267, 275, 283, 291, 299, 307, 315, 323, 331, 339, 347, 355, 363, 379 \\ 
28 & 3, 11, 19, 27, 35, 43, 51, 59, 67, 75, 83, 91, 99, 107, 115, 123, 131, 139, 147, 155, 163, 171, 187 \\ 
32 & 370, 378, 386, 394, 402, 410, 418, 426, 434, 442, 450, 458, 466, 474, 482, 490, 498, 506 \\ 
33 & 178, 186, 194, 202, 210, 218, 226, 234, 242, 250, 258, 266, 274, 282, 290, 298, 306, 314, 322, 330, 338, 346, 354, 362 \\ 
34 & 2, 10, 18, 26, 34, 42, 50, 58, 66, 74, 82, 90, 98, 106, 114, 122, 130, 138, 146, 154, 162, 170 \\ 
38 & 369, 377, 385, 393, 401, 409, 417, 425, 433, 441, 449, 457, 465, 473, 481, 489, 497, 505 \\ 
39 & 185, 193, 201, 209, 217, 225, 233, 241, 249, 257, 265, 273, 281, 289, 297, 305, 313, 321, 329, 337, 345, 353, 361 \\ 
40 & 1, 9, 17, 25, 33, 41, 49, 57, 65, 73, 81, 89, 97, 105, 113, 121, 129, 137, 145, 153, 161, 169, 177 \\ 
44 & 368, 376, 384, 392, 400, 408, 416, 424, 432, 440, 448, 456, 464, 472, 480, 488, 496, 504 \\ 
45 & 192, 200, 208, 216, 224, 232, 240, 248, 256, 264, 272, 280, 288, 296, 304, 312, 320, 328, 336, 344, 352, 360 \\ 
46 & 0, 8, 16, 24, 32, 40, 48, 56, 64, 72, 80, 88, 96, 104, 112, 120, 128, 136, 144, 152, 160, 168, 176, 184 \\ 
\end{tabular}
}
}
}
\captionof{table}{Localization of the key in the MSP430's SRAM for the sections shown in Fig.~\ref{fig:msp_sections_localization}.\label{tab:msp_localization}}
\end{minipage}

\end{document}

%% file: figs/background/SRAM_new_TLS.pdf_tex
\begingroup%
  \makeatletter%
  \providecommand\color[2][]{%
    \errmessage{(Inkscape) Color is used for the text in Inkscape, but the package 'color.sty' is not loaded}%
    \renewcommand\color[2][]{}%
  }%
  \providecommand\transparent[1]{%
    \errmessage{(Inkscape) Transparency is used (non-zero) for the text in Inkscape, but the package 'transparent.sty' is not loaded}%
    \renewcommand\transparent[1]{}%
  }%
  \providecommand\rotatebox[2]{#2}%
  \newcommand*\fsize{\dimexpr\f@size pt\relax}%
  \newcommand*\lineheight[1]{\fontsize{\fsize}{#1\fsize}\selectfont}%
  \ifx\svgwidth\undefined%
    \setlength{\unitlength}{161.25bp}%
    \ifx\svgscale\undefined%
      \relax%
    \else%
      \setlength{\unitlength}{\unitlength * \real{\svgscale}}%
    \fi%
  \else%
    \setlength{\unitlength}{\svgwidth}%
  \fi%
  \global\let\svgwidth\undefined%
  \global\let\svgscale\undefined%
  \makeatother%
  \begin{picture}(1,1.4786174)%
    \lineheight{1}%
    \setlength\tabcolsep{0pt}%
    \put(0,0){\includegraphics[width=\unitlength,page=1]{SRAM_new_TLS.pdf}}%
    \put(0.81801111,1.17073596){\color[rgb]{0,0,0}\makebox(0,0)[lt]{\lineheight{1.25}\smash{\begin{tabular}[t]{l}M\textsubscript{p2}\end{tabular}}}}%
    \put(0,0){\includegraphics[width=\unitlength,page=2]{SRAM_new_TLS.pdf}}%
    \put(0.82043823,0.72484941){\color[rgb]{0,0,0}\makebox(0,0)[lt]{\lineheight{1.25}\smash{\begin{tabular}[t]{l}M\textsubscript{n2}\end{tabular}}}}%
    \put(0,0){\includegraphics[width=\unitlength,page=3]{SRAM_new_TLS.pdf}}%
    \put(0.07123654,0.7236334){\color[rgb]{0,0,0}\makebox(0,0)[lt]{\lineheight{1.25}\smash{\begin{tabular}[t]{l}M\textsubscript{n1}\end{tabular}}}}%
    \put(0,0){\includegraphics[width=\unitlength,page=4]{SRAM_new_TLS.pdf}}%
    \put(0.26359684,0.35751634){\color[rgb]{0,0,0}\makebox(0,0)[lt]{\lineheight{1.25}\smash{\begin{tabular}[t]{l}TLS-sensitive areas\end{tabular}}}}%
    \put(0,0){\includegraphics[width=\unitlength,page=5]{SRAM_new_TLS.pdf}}%
    \put(0.07316659,1.17644438){\color[rgb]{0,0,0}\makebox(0,0)[lt]{\lineheight{1.25}\smash{\begin{tabular}[t]{l}M\textsubscript{p1}\end{tabular}}}}%
    \put(0,0){\includegraphics[width=\unitlength,page=6]{SRAM_new_TLS.pdf}}%
    \put(0.03175183,0.87950085){\color[rgb]{0,0,0}\makebox(0,0)[lt]{\lineheight{1.25}\smash{\begin{tabular}[t]{l}\tiny{GND$+$}\end{tabular}}}}%
    \put(0,0){\includegraphics[width=\unitlength,page=7]{SRAM_new_TLS.pdf}}%
    \put(0.26494299,0.22276406){\color[rgb]{0,0,0}\makebox(0,0)[lt]{\lineheight{1.25}\smash{\begin{tabular}[t]{l}Increased leakage current\end{tabular}}}}%
    \put(0,0){\includegraphics[width=\unitlength,page=8]{SRAM_new_TLS.pdf}}%
    \put(0.81040117,1.0722734){\color[rgb]{0,0,0}\makebox(0,0)[lt]{\lineheight{1.25}\smash{\begin{tabular}[t]{l}\tiny{VCC$-$}\end{tabular}}}}%
    \put(0.81165014,1.02361377){\makebox(0,0)[lt]{\lineheight{1.25}\smash{\begin{tabular}[t]{l}\tiny{V\textsubscript{Seeb.}}\end{tabular}}}}%
    \put(0,0){\includegraphics[width=\unitlength,page=9]{SRAM_new_TLS.pdf}}%
    \put(0.25879926,0.07606836){\color[rgb]{0,0,0}\makebox(0,0)[lt]{\lineheight{1.25}\smash{\begin{tabular}[t]{l}Low/High-ohmic channel\end{tabular}}}}%
    \put(0,0){\includegraphics[width=\unitlength,page=10]{SRAM_new_TLS.pdf}}%
    \put(0.58976744,0.49489647){\color[rgb]{0,0,0}\makebox(0,0)[t]{\lineheight{1.25}\smash{\begin{tabular}[t]{c}GND\end{tabular}}}}%
    \put(0,0){\includegraphics[width=\unitlength,page=11]{SRAM_new_TLS.pdf}}%
    \put(0.58653023,1.40811615){\color[rgb]{0,0,0}\makebox(0,0)[t]{\lineheight{1.25}\smash{\begin{tabular}[t]{c}VCC\end{tabular}}}}%
    \put(0,0){\includegraphics[width=\unitlength,page=12]{SRAM_new_TLS.pdf}}%
    \put(0.03224251,0.83062858){\makebox(0,0)[lt]{\lineheight{1.25}\smash{\begin{tabular}[t]{l}\tiny{V\textsubscript{Seeb.}}\end{tabular}}}}%
    \put(0,0){\includegraphics[width=\unitlength,page=13]{SRAM_new_TLS.pdf}}%
  \end{picture}%
\endgroup%

%% file: figs/background/SRAM_new_LLSI.pdf_tex
\begingroup%
  \makeatletter%
  \providecommand\color[2][]{%
    \errmessage{(Inkscape) Color is used for the text in Inkscape, but the package 'color.sty' is not loaded}%
    \renewcommand\color[2][]{}%
  }%
  \providecommand\transparent[1]{%
    \errmessage{(Inkscape) Transparency is used (non-zero) for the text in Inkscape, but the package 'transparent.sty' is not loaded}%
    \renewcommand\transparent[1]{}%
  }%
  \providecommand\rotatebox[2]{#2}%
  \newcommand*\fsize{\dimexpr\f@size pt\relax}%
  \newcommand*\lineheight[1]{\fontsize{\fsize}{#1\fsize}\selectfont}%
  \ifx\svgwidth\undefined%
    \setlength{\unitlength}{161.25bp}%
    \ifx\svgscale\undefined%
      \relax%
    \else%
      \setlength{\unitlength}{\unitlength * \real{\svgscale}}%
    \fi%
  \else%
    \setlength{\unitlength}{\svgwidth}%
  \fi%
  \global\let\svgwidth\undefined%
  \global\let\svgscale\undefined%
  \makeatother%
  \begin{picture}(1,1.4786174)%
    \lineheight{1}%
    \setlength\tabcolsep{0pt}%
    \put(0,0){\includegraphics[width=\unitlength,page=1]{SRAM_new_LLSI.pdf}}%
    \put(0.81801111,1.17073596){\color[rgb]{0,0,0}\makebox(0,0)[lt]{\lineheight{1.25}\smash{\begin{tabular}[t]{l}M\textsubscript{p2}\end{tabular}}}}%
    \put(0,0){\includegraphics[width=\unitlength,page=2]{SRAM_new_LLSI.pdf}}%
    \put(0.82043823,0.72484941){\color[rgb]{0,0,0}\makebox(0,0)[lt]{\lineheight{1.25}\smash{\begin{tabular}[t]{l}M\textsubscript{n2}\end{tabular}}}}%
    \put(0,0){\includegraphics[width=\unitlength,page=3]{SRAM_new_LLSI.pdf}}%
    \put(0.07123654,0.7236334){\color[rgb]{0,0,0}\makebox(0,0)[lt]{\lineheight{1.25}\smash{\begin{tabular}[t]{l}M\textsubscript{n1}\end{tabular}}}}%
    \put(0,0){\includegraphics[width=\unitlength,page=4]{SRAM_new_LLSI.pdf}}%
    \put(0.07316659,1.17644438){\color[rgb]{0,0,0}\makebox(0,0)[lt]{\lineheight{1.25}\smash{\begin{tabular}[t]{l}M\textsubscript{p1}\end{tabular}}}}%
    \put(0,0){\includegraphics[width=\unitlength,page=5]{SRAM_new_LLSI.pdf}}%
    \put(0.25879926,0.07606836){\color[rgb]{0,0,0}\makebox(0,0)[lt]{\lineheight{1.25}\smash{\begin{tabular}[t]{l}Low/High-ohmic channel\end{tabular}}}}%
    \put(0,0){\includegraphics[width=\unitlength,page=6]{SRAM_new_LLSI.pdf}}%
    \put(0.58976744,0.49489647){\color[rgb]{0,0,0}\makebox(0,0)[t]{\lineheight{1.25}\smash{\begin{tabular}[t]{c}GND\end{tabular}}}}%
    \put(0,0){\includegraphics[width=\unitlength,page=7]{SRAM_new_LLSI.pdf}}%
    \put(0.58653023,1.40811615){\color[rgb]{0,0,0}\makebox(0,0)[t]{\lineheight{1.25}\smash{\begin{tabular}[t]{c}VCC\end{tabular}}}}%
    \put(0,0){\includegraphics[width=\unitlength,page=8]{SRAM_new_LLSI.pdf}}%
    \put(0.25532663,0.346645){\color[rgb]{0,0,0}\makebox(0,0)[lt]{\lineheight{1.25}\smash{\begin{tabular}[t]{l}LLSI signal origin\end{tabular}}}}%
    \put(0,0){\includegraphics[width=\unitlength,page=9]{SRAM_new_LLSI.pdf}}%
    \put(0.25776563,0.2162399){\color[rgb]{0,0,0}\makebox(0,0)[lt]{\lineheight{1.25}\smash{\begin{tabular}[t]{l}Power supply modulation\end{tabular}}}}%
    \put(0,0){\includegraphics[width=\unitlength,page=10]{SRAM_new_LLSI.pdf}}%
  \end{picture}%
\endgroup%

%% file: figs/results/ultrascale/single_bit_acc.tex
\begin{tikzpicture}
	
\begin{axis}[
axis background/.style={BgStyle},
axis line style={white},
height=\axheight,
legend cell align={left},
legend style={draw opacity=1, draw=none, text opacity=1 , fill=white, fill opacity=0.9, font=\scriptsize},
legend pos=south east,
tick align=outside,
tick pos=left,
ticklabel style = {font=\scriptsize},
width=\axwidth,
x grid style={white},
xlabel={Epoch},
xmajorgrids,
xmin=0, xmax=40,
xtick style={color=white!33.3333333333333!black},
y grid style={white},
ylabel={Accuracy},
ymajorgrids,
ymin=0.4, ymax=1.05,
ytick style={color=white!33.3333333333333!black},
label style = {font=\footnotesize}
]
\addplot [thick, color0]
table {%
	1 0.505050480365753
	2 0.51262629032135
	3 0.522727251052856
	4 0.527777791023254
	5 0.537878811359406
	6 0.595959603786469
	7 0.65656566619873
	8 0.69696968793869
	9 0.833333313465118
	10 0.911616146564484
	11 0.967171728610992
	12 0.974747478961945
	13 0.982323229312897
	14 0.994949519634247
	15 0.980000019073486
	16 0.964646458625793
	17 0.989898979663849
	18 0.98737370967865
	19 0.997474730014801
	20 1
	21 0.98737370967865
	22 0.994949519634247
	23 1
	24 0.989898979663849
	25 0.997474730014801
	26 0.994949519634247
	27 0.989898979663849
	28 0.997474730014801
	29 0.98737370967865
	30 1
	31 0.984848499298096
	32 0.997474730014801
	33 0.997474730014801
	34 0.98737370967865
	35 0.997474730014801
	36 0.997474730014801
	37 0.994949519634247
	38 0.992424249649048
	39 1
	40 1
};
\addlegendentry{Training}
\addplot [thick, color2]
table {%
	1 0.512499988079071
	2 0.5
	3 0.5
	4 0.5
	5 0.537500023841858
	6 0.5
	7 0.5625
	8 0.737500011920929
	9 0.850000023841858
	10 0.925000011920929
	11 0.925000011920929
	12 0.987500011920929
	13 0.949999988079071
	14 1
	15 0.987500011920929
	16 1
	17 1
	18 1
	19 1
	20 1
	21 1
	22 1
	23 0.987500011920929
	24 1
	25 1
	26 0.987500011920929
	27 0.987500011920929
	28 1
	29 1
	30 1
	31 1
	32 1
	33 1
	34 1
	35 1
	36 1
	37 1
	38 1
	39 1
	40 1
};
\addlegendentry{Validation}
\end{axis}

\end{tikzpicture}

%% file: figs/results/ultrascale/bits_in_parallel_test_acc_4_bits_up_to_256.tex
\begin{tikzpicture}

\begin{axis}[
axis background/.style={BgStyle},
axis line style={white},
height=\axheight,
label style = {font=\footnotesize},
legend cell align={left},
legend style={draw opacity=1, draw=none, text opacity=1 , fill=white, fill opacity=0.6, font=\scriptsize},
legend pos=south west,
tick align=outside,
tick pos=both,
ticklabel style = {font=\scriptsize},
width=\axwidth,
x grid style={white},
xmajorgrids,
xmin=0, xmax=260,
xtick style={color=white!33.3333333333333!black},
y grid style={white},
ymajorgrids,
ymin=0.55, ymax=1.01,
ytick style={color=white!33.3333333333333!black},
xtick={8,16,32,64,128,256},
ylabel={Test accuracy},
xlabel={Number of bits learned in one network},
]

\addplot [thick, color0]
table {%
	0 0
};
\addlegendentry{Full image}

\addplot [thick, color1]
table {%
	0 0
};
\addlegendentry{128$\times$128\,px}

\addplot [thick, color2]
table {%
	0 0
};
\addlegendentry{64$\times$64\,px}

\addplot [thick, color0, mark=*, mark size=1, mark options={MarkStyle}]
table {%
2 1
4 1
8 1
16 1
32 0.977011494252874
64 0.793103448275862
128 0.586206896551724
256 0.574712643678161
};

\addplot [thick, color1, mark=*, mark size=1, mark options={MarkStyle}]
table {%
2 1
4 1
8 1
16 1
32 1
64 1
128 0.942528735632184
256 0.666666666666667
};

\addplot [thick, color2, mark=*, mark size=1, mark options={MarkStyle}]
table {%
2 1
4 1
8 1
16 1
32 1
64 1
128 0.977011494252874
256 0.919540229885057
};

\addplot [thick, color0, mark=triangle*, mark size=1, mark options={MarkStyle,rotate=180}]
table {%
2 1
4 1
8 1
16 1
32 0.988505747126437
64 0.747126436781609
128 0.655172413793103
256 0.505747126436782
};

\addplot [thick, color1, mark=triangle*, mark size=1, mark options={MarkStyle,rotate=180}]
table {%
2 1
4 1
8 1
16 1
32 1
64 0.977011494252874
128 0.96551724137931
256 0.735632183908046
};

\addplot [thick, color2, mark=triangle*, mark size=1, mark options={MarkStyle,rotate=180}]
table {%
2 1
4 1
8 1
16 1
32 1
64 0.988505747126437
128 0.96551724137931
256 0.885057471264368
};

\addplot [thick, color0, mark=square*, mark size=1, mark options={MarkStyle}]
table {%
4 1
8 1
16 1
32 1
64 0.758620689655172
128 0.632183908045977
256 0.597701149425287
};

\addplot [thick, color1, mark=square*, mark size=1, mark options={MarkStyle}]
table {%
4 1
8 1
16 1
32 1
64 0.988505747126437
128 0.896551724137931
256 0.758620689655172
};

\addplot [thick, color2, mark=square*, mark size=1, mark options={MarkStyle}]
table {%
4 1
8 1
16 1
32 1
64 1
128 1
256 0.827586206896552
};

\addplot [thick, color0, mark=diamond*, mark size=1, mark options={MarkStyle}]
table {%
4 1
8 1
16 1
32 0.988505747126437
64 0.850574712643678
128 0.597701149425287
256 0.597701149425287
};

\addplot [thick, color1, mark=diamond*, mark size=1, mark options={MarkStyle}]
table {%
4 1
8 1
16 1
32 1
64 1
128 0.931034482758621
256 0.563218390804598
};

\addplot [thick, color2, mark=diamond*, mark size=1, mark options={MarkStyle}]
table {%
4 1
8 1
16 1
32 1
64 0.988505747126437
128 0.988505747126437
256 0.862068965517241
};

\end{axis}

\end{tikzpicture}

%% file: figs/results/ultrascale/single_bits_num_imgs.tex
\begin{tikzpicture}
	
	\begin{axis}[
		axis background/.style={BgStyle},
		axis line style={white},
		height=\axheight,
		legend cell align={left},
		legend style={draw opacity=1, draw=none, text opacity=1 , fill=white, fill opacity=0.6, font=\scriptsize},
		legend pos=south east,
		tick align=outside,
		tick pos=both,
		ticklabel style = {font=\scriptsize},
		width=\axwidth,
		x grid style={white},
		xmajorgrids,
		xmin=5, xmax=401,
		xtick style={color=white!33.3333333333333!black},
		xtick={50,100,150,200,250,300,350,400},
		xlabel={Number of images used for training},
		y grid style={white},
		ymajorgrids,
		ymin=0.5, ymax=1.01,
		ytick style={color=white!33.3333333333333!black},
		ylabel={Test accuracy},
		label style = {font=\footnotesize}
		]
		\addplot [thick, color0]
		table {%
			410 1
		};
		\addlegendentry{Full image}
		\addplot [thick, color1]
		table {%
			410 1
		};
		\addlegendentry{128$\times$128\,px}
		\addplot [thick, color2]
		table {%
			410 1
		};
		\addlegendentry{64$\times$64\,px}
		
		\addplot [thick, black, mark=*, mark size=1, mark options={MarkStyle}]
		table {%
			410 1
		};
		\addlegendentry{Bit 0}
		\addplot [thick, black, mark=triangle*, mark size=1, mark options={MarkStyle,rotate=180}]
		table {%
			410 1
		};
		\addlegendentry{Bit 1}
		\addplot [thick, black, mark=square*, mark size=1, mark options={MarkStyle}]
		table {%
			410 1
		};
		\addlegendentry{Bit 2}

		\addplot [thick, color0, mark=*, mark size=1, mark options={MarkStyle}]
		table {%
			400 1
			300 1
			200 1
			150 0.983644859813084
			100 0.98744769874477
			75 0.783730158730159
			50 0.636363636363636
			40 0.615241635687732
			30 0.594890510948905
			20 0.514336917562724
			10 0.507042253521127
		};

		\addplot [thick, color1, mark=*, mark size=1, mark options={MarkStyle, draw=black, thin}]
		table {%
			400 1
			300 1
			200 1
			150 1
			100 1
			75 1
			50 0.958333333333333
			40 0.925650557620818
			30 0.883211678832117
			20 0.840501792114695
			10 0.688380281690141
		};

		\addplot [thick, color2, mark=*, mark size=1, mark options={MarkStyle}]
		table {%
			400 1
			300 1
			200 1
			150 1
			100 1
			75 1
			50 0.998106060606061
			40 0.981412639405205
			30 0.936131386861314
			20 0.845878136200717
			10 0.711267605633803
		};

		\addplot [thick, color0, mark=triangle*, mark size=1, mark options={MarkStyle,rotate=180}]
		table {%
			400 1
			300 0.996415770609319
			200 0.981481481481482
			150 0.869158878504673
			100 0.780334728033473
			75 0.755952380952381
			50 0.691287878787879
			40 0.656133828996283
			30 0.589416058394161
			20 0.551971326164875
			10 0.554577464788732
		};

		\addplot [thick, color1, mark=triangle*, mark size=1, mark options={MarkStyle,rotate=180}]
		table {%
			400 1
			300 1
			200 1
			150 1
			100 1
			75 0.982142857142857
			50 0.975378787878788
			40 0.942379182156134
			30 0.832116788321168
			20 0.770609318996416
			10 0.60387323943662
		};

		\addplot [thick, color2, mark=triangle*, mark size=1, mark options={MarkStyle,rotate=180}]
		table {%
			400 1
			300 1
			200 1
			150 1
			100 1
			75 1
			50 1
			40 0.996282527881041
			30 0.987226277372263
			20 0.935483870967742
			10 0.665492957746479
		};

		\addplot [thick, color0, mark=square*, mark size=1, mark options={MarkStyle}]
		table {%
			400 1
			300 1
			200 0.952380952380952
			150 0.911214953271028
			100 0.730125523012552
			75 0.692460317460317
			50 0.615530303030303
			40 0.617100371747212
			30 0.563868613138686
			20 0.521505376344086
			10 0.501760563380282
		};

		\addplot [thick, color1, mark=square*, mark size=1, mark options={MarkStyle}]
		table {%
			400 1
			300 1
			200 1
			150 1
			100 1
			75 0.986111111111111
			50 0.965909090909091
			40 0.91635687732342
			30 0.686131386861314
			20 0.763440860215054
			10 0.568661971830986
		};

		\addplot [thick, color2, mark=square*, mark size=1, mark options={MarkStyle}]
		table {%
			400 1
			300 1
			200 1
			150 1
			100 1
			75 1
			50 0.998106060606061
			40 0.983271375464684
			30 0.932481751824818
			20 0.876344086021505
			10 0.663732394366197
		};

	\end{axis}
	
\end{tikzpicture}

%% file: figs/results/ultrascale/bits_parallel_few_imgs.tex
\begin{tikzpicture}

\begin{axis}[
axis background/.style={BgStyle},
axis line style={white},
height=\axheight,
label style = {font=\footnotesize},
legend style={draw opacity=1, draw=none, text opacity=1 , fill=white, fill opacity=0.9, font=\scriptsize},
legend pos=south east,
tick align=outside,
tick pos=both,
ticklabel style = {font=\scriptsize},
width=\axwidth,
x grid style={white},
xmajorgrids,
xmin=-0.5, xmax=127.5,
xtick style={color=white!33.3333333333333!black},
xticklabel style = {rotate=90.0},
y grid style={white},
ymajorgrids,
ymin=0, ymax=1,
ytick style={color=white!33.3333333333333!black},
xlabel={Bit position in the key},
ylabel={Test accuracy},
xtick={0, 16, 32, 48, 64, 80, 96, 112, 127}
]
\draw[draw=color2,fill=color2] (axis cs:-0.25,0) rectangle (axis cs:0.25,0.956521739130435);
\addlegendimage{ybar,ybar legend,draw=white!93.3333333333333!black,fill=color2};
\addlegendentry{Accuracy $>$ 0.75}
\draw[draw=color2,fill=color2] (axis cs:0.75,0) rectangle (axis cs:1.25,0.869565217391304);
\draw[draw=color2,fill=color2] (axis cs:1.75,0) rectangle (axis cs:2.25,0.869565217391304);
\draw[draw=color2,fill=color2] (axis cs:2.75,0) rectangle (axis cs:3.25,0.956521739130435);
\draw[draw=color2,fill=color2] (axis cs:3.75,0) rectangle (axis cs:4.25,0.91304347826087);
\draw[draw=color2,fill=color2] (axis cs:4.75,0) rectangle (axis cs:5.25,1);
\draw[draw=color0,fill=color0] (axis cs:5.75,0) rectangle (axis cs:6.25,0.652173913043478);
\addlegendimage{ybar,ybar legend,draw=white!93.3333333333333!black,fill=color0};
\addlegendentry{Accuracy $\leq$ 0.75}
\draw[draw=color0,fill=color0] (axis cs:6.75,0) rectangle (axis cs:7.25,0.434782608695652);
\draw[draw=color0,fill=color0] (axis cs:7.75,0) rectangle (axis cs:8.25,0.521739130434783);
\draw[draw=color0,fill=color0] (axis cs:8.75,0) rectangle (axis cs:9.25,0.565217391304348);
\draw[draw=color0,fill=color0] (axis cs:9.75,0) rectangle (axis cs:10.25,0.478260869565217);
\draw[draw=color0,fill=color0] (axis cs:10.75,0) rectangle (axis cs:11.25,0.521739130434783);
\draw[draw=color0,fill=color0] (axis cs:11.75,0) rectangle (axis cs:12.25,0.391304347826087);
\draw[draw=color0,fill=color0] (axis cs:12.75,0) rectangle (axis cs:13.25,0.521739130434783);
\draw[draw=color0,fill=color0] (axis cs:13.75,0) rectangle (axis cs:14.25,0.521739130434783);
\draw[draw=color0,fill=color0] (axis cs:14.75,0) rectangle (axis cs:15.25,0.565217391304348);
\draw[draw=color0,fill=color0] (axis cs:15.75,0) rectangle (axis cs:16.25,0.565217391304348);
\draw[draw=color0,fill=color0] (axis cs:16.75,0) rectangle (axis cs:17.25,0.304347826086957);
\draw[draw=color0,fill=color0] (axis cs:17.75,0) rectangle (axis cs:18.25,0.565217391304348);
\draw[draw=color0,fill=color0] (axis cs:18.75,0) rectangle (axis cs:19.25,0.652173913043478);
\draw[draw=color0,fill=color0] (axis cs:19.75,0) rectangle (axis cs:20.25,0.478260869565217);
\draw[draw=color0,fill=color0] (axis cs:20.75,0) rectangle (axis cs:21.25,0.565217391304348);
\draw[draw=color0,fill=color0] (axis cs:21.75,0) rectangle (axis cs:22.25,0.565217391304348);
\draw[draw=color0,fill=color0] (axis cs:22.75,0) rectangle (axis cs:23.25,0.521739130434783);
\draw[draw=color0,fill=color0] (axis cs:23.75,0) rectangle (axis cs:24.25,0.434782608695652);
\draw[draw=color0,fill=color0] (axis cs:24.75,0) rectangle (axis cs:25.25,0.521739130434783);
\draw[draw=color0,fill=color0] (axis cs:25.75,0) rectangle (axis cs:26.25,0.521739130434783);
\draw[draw=color0,fill=color0] (axis cs:26.75,0) rectangle (axis cs:27.25,0.565217391304348);
\draw[draw=color0,fill=color0] (axis cs:27.75,0) rectangle (axis cs:28.25,0.434782608695652);
\draw[draw=color0,fill=color0] (axis cs:28.75,0) rectangle (axis cs:29.25,0.478260869565217);
\draw[draw=color0,fill=color0] (axis cs:29.75,0) rectangle (axis cs:30.25,0.521739130434783);
\draw[draw=color0,fill=color0] (axis cs:30.75,0) rectangle (axis cs:31.25,0.434782608695652);
\draw[draw=color2,fill=color2] (axis cs:31.75,0) rectangle (axis cs:32.25,0.869565217391304);
\draw[draw=color2,fill=color2] (axis cs:32.75,0) rectangle (axis cs:33.25,0.869565217391304);
\draw[draw=color2,fill=color2] (axis cs:33.75,0) rectangle (axis cs:34.25,0.782608695652174);
\draw[draw=color2,fill=color2] (axis cs:34.75,0) rectangle (axis cs:35.25,0.869565217391304);
\draw[draw=color2,fill=color2] (axis cs:35.75,0) rectangle (axis cs:36.25,0.91304347826087);
\draw[draw=color2,fill=color2] (axis cs:36.75,0) rectangle (axis cs:37.25,0.91304347826087);
\draw[draw=color0,fill=color0] (axis cs:37.75,0) rectangle (axis cs:38.25,0.608695652173913);
\draw[draw=color0,fill=color0] (axis cs:38.75,0) rectangle (axis cs:39.25,0.478260869565217);
\draw[draw=color0,fill=color0] (axis cs:39.75,0) rectangle (axis cs:40.25,0.695652173913043);
\draw[draw=color0,fill=color0] (axis cs:40.75,0) rectangle (axis cs:41.25,0.304347826086957);
\draw[draw=color0,fill=color0] (axis cs:41.75,0) rectangle (axis cs:42.25,0.434782608695652);
\draw[draw=color0,fill=color0] (axis cs:42.75,0) rectangle (axis cs:43.25,0.434782608695652);
\draw[draw=color0,fill=color0] (axis cs:43.75,0) rectangle (axis cs:44.25,0.521739130434783);
\draw[draw=color0,fill=color0] (axis cs:44.75,0) rectangle (axis cs:45.25,0.608695652173913);
\draw[draw=color0,fill=color0] (axis cs:45.75,0) rectangle (axis cs:46.25,0.478260869565217);
\draw[draw=color0,fill=color0] (axis cs:46.75,0) rectangle (axis cs:47.25,0.565217391304348);
\draw[draw=color0,fill=color0] (axis cs:47.75,0) rectangle (axis cs:48.25,0.521739130434783);
\draw[draw=color0,fill=color0] (axis cs:48.75,0) rectangle (axis cs:49.25,0.521739130434783);
\draw[draw=color0,fill=color0] (axis cs:49.75,0) rectangle (axis cs:50.25,0.478260869565217);
\draw[draw=color0,fill=color0] (axis cs:50.75,0) rectangle (axis cs:51.25,0.608695652173913);
\draw[draw=color0,fill=color0] (axis cs:51.75,0) rectangle (axis cs:52.25,0.608695652173913);
\draw[draw=color0,fill=color0] (axis cs:52.75,0) rectangle (axis cs:53.25,0.521739130434783);
\draw[draw=color0,fill=color0] (axis cs:53.75,0) rectangle (axis cs:54.25,0.608695652173913);
\draw[draw=color0,fill=color0] (axis cs:54.75,0) rectangle (axis cs:55.25,0.478260869565217);
\draw[draw=color0,fill=color0] (axis cs:55.75,0) rectangle (axis cs:56.25,0.521739130434783);
\draw[draw=color0,fill=color0] (axis cs:56.75,0) rectangle (axis cs:57.25,0.347826086956522);
\draw[draw=color0,fill=color0] (axis cs:57.75,0) rectangle (axis cs:58.25,0.521739130434783);
\draw[draw=color0,fill=color0] (axis cs:58.75,0) rectangle (axis cs:59.25,0.521739130434783);
\draw[draw=color0,fill=color0] (axis cs:59.75,0) rectangle (axis cs:60.25,0.304347826086957);
\draw[draw=color0,fill=color0] (axis cs:60.75,0) rectangle (axis cs:61.25,0.391304347826087);
\draw[draw=color0,fill=color0] (axis cs:61.75,0) rectangle (axis cs:62.25,0.565217391304348);
\draw[draw=color0,fill=color0] (axis cs:62.75,0) rectangle (axis cs:63.25,0.434782608695652);
\draw[draw=color2,fill=color2] (axis cs:63.75,0) rectangle (axis cs:64.25,1);
\draw[draw=color2,fill=color2] (axis cs:64.75,0) rectangle (axis cs:65.25,0.826086956521739);
\draw[draw=color2,fill=color2] (axis cs:65.75,0) rectangle (axis cs:66.25,0.91304347826087);
\draw[draw=color2,fill=color2] (axis cs:66.75,0) rectangle (axis cs:67.25,0.91304347826087);
\draw[draw=color2,fill=color2] (axis cs:67.75,0) rectangle (axis cs:68.25,0.91304347826087);
\draw[draw=color2,fill=color2] (axis cs:68.75,0) rectangle (axis cs:69.25,0.91304347826087);
\draw[draw=color0,fill=color0] (axis cs:69.75,0) rectangle (axis cs:70.25,0.695652173913043);
\draw[draw=color0,fill=color0] (axis cs:70.75,0) rectangle (axis cs:71.25,0.391304347826087);
\draw[draw=color0,fill=color0] (axis cs:71.75,0) rectangle (axis cs:72.25,0.565217391304348);
\draw[draw=color0,fill=color0] (axis cs:72.75,0) rectangle (axis cs:73.25,0.652173913043478);
\draw[draw=color0,fill=color0] (axis cs:73.75,0) rectangle (axis cs:74.25,0.565217391304348);
\draw[draw=color0,fill=color0] (axis cs:74.75,0) rectangle (axis cs:75.25,0.434782608695652);
\draw[draw=color0,fill=color0] (axis cs:75.75,0) rectangle (axis cs:76.25,0.478260869565217);
\draw[draw=color0,fill=color0] (axis cs:76.75,0) rectangle (axis cs:77.25,0.478260869565217);
\draw[draw=color0,fill=color0] (axis cs:77.75,0) rectangle (axis cs:78.25,0.521739130434783);
\draw[draw=color0,fill=color0] (axis cs:78.75,0) rectangle (axis cs:79.25,0.521739130434783);
\draw[draw=color0,fill=color0] (axis cs:79.75,0) rectangle (axis cs:80.25,0.739130434782609);
\draw[draw=color0,fill=color0] (axis cs:80.75,0) rectangle (axis cs:81.25,0.652173913043478);
\draw[draw=color0,fill=color0] (axis cs:81.75,0) rectangle (axis cs:82.25,0.478260869565217);
\draw[draw=color0,fill=color0] (axis cs:82.75,0) rectangle (axis cs:83.25,0.478260869565217);
\draw[draw=color0,fill=color0] (axis cs:83.75,0) rectangle (axis cs:84.25,0.695652173913043);
\draw[draw=color0,fill=color0] (axis cs:84.75,0) rectangle (axis cs:85.25,0.521739130434783);
\draw[draw=color0,fill=color0] (axis cs:85.75,0) rectangle (axis cs:86.25,0.565217391304348);
\draw[draw=color0,fill=color0] (axis cs:86.75,0) rectangle (axis cs:87.25,0.478260869565217);
\draw[draw=color0,fill=color0] (axis cs:87.75,0) rectangle (axis cs:88.25,0.521739130434783);
\draw[draw=color0,fill=color0] (axis cs:88.75,0) rectangle (axis cs:89.25,0.521739130434783);
\draw[draw=color0,fill=color0] (axis cs:89.75,0) rectangle (axis cs:90.25,0.521739130434783);
\draw[draw=color0,fill=color0] (axis cs:90.75,0) rectangle (axis cs:91.25,0.608695652173913);
\draw[draw=color0,fill=color0] (axis cs:91.75,0) rectangle (axis cs:92.25,0.478260869565217);
\draw[draw=color0,fill=color0] (axis cs:92.75,0) rectangle (axis cs:93.25,0.478260869565217);
\draw[draw=color0,fill=color0] (axis cs:93.75,0) rectangle (axis cs:94.25,0.608695652173913);
\draw[draw=color0,fill=color0] (axis cs:94.75,0) rectangle (axis cs:95.25,0.304347826086957);
\draw[draw=color2,fill=color2] (axis cs:95.75,0) rectangle (axis cs:96.25,1);
\draw[draw=color2,fill=color2] (axis cs:96.75,0) rectangle (axis cs:97.25,0.91304347826087);
\draw[draw=color2,fill=color2] (axis cs:97.75,0) rectangle (axis cs:98.25,1);
\draw[draw=color2,fill=color2] (axis cs:98.75,0) rectangle (axis cs:99.25,0.91304347826087);
\draw[draw=color2,fill=color2] (axis cs:99.75,0) rectangle (axis cs:100.25,0.869565217391304);
\draw[draw=color2,fill=color2] (axis cs:100.75,0) rectangle (axis cs:101.25,0.956521739130435);
\draw[draw=color0,fill=color0] (axis cs:101.75,0) rectangle (axis cs:102.25,0.434782608695652);
\draw[draw=color0,fill=color0] (axis cs:102.75,0) rectangle (axis cs:103.25,0.391304347826087);
\draw[draw=color0,fill=color0] (axis cs:103.75,0) rectangle (axis cs:104.25,0.217391304347826);
\draw[draw=color0,fill=color0] (axis cs:104.75,0) rectangle (axis cs:105.25,0.434782608695652);
\draw[draw=color0,fill=color0] (axis cs:105.75,0) rectangle (axis cs:106.25,0.608695652173913);
\draw[draw=color0,fill=color0] (axis cs:106.75,0) rectangle (axis cs:107.25,0.608695652173913);
\draw[draw=color0,fill=color0] (axis cs:107.75,0) rectangle (axis cs:108.25,0.565217391304348);
\draw[draw=color0,fill=color0] (axis cs:108.75,0) rectangle (axis cs:109.25,0.478260869565217);
\draw[draw=color0,fill=color0] (axis cs:109.75,0) rectangle (axis cs:110.25,0.391304347826087);
\draw[draw=color0,fill=color0] (axis cs:110.75,0) rectangle (axis cs:111.25,0.521739130434783);
\draw[draw=color0,fill=color0] (axis cs:111.75,0) rectangle (axis cs:112.25,0.521739130434783);
\draw[draw=color0,fill=color0] (axis cs:112.75,0) rectangle (axis cs:113.25,0.478260869565217);
\draw[draw=color0,fill=color0] (axis cs:113.75,0) rectangle (axis cs:114.25,0.652173913043478);
\draw[draw=color0,fill=color0] (axis cs:114.75,0) rectangle (axis cs:115.25,0.521739130434783);
\draw[draw=color0,fill=color0] (axis cs:115.75,0) rectangle (axis cs:116.25,0.478260869565217);
\draw[draw=color0,fill=color0] (axis cs:116.75,0) rectangle (axis cs:117.25,0.521739130434783);
\draw[draw=color0,fill=color0] (axis cs:117.75,0) rectangle (axis cs:118.25,0.434782608695652);
\draw[draw=color0,fill=color0] (axis cs:118.75,0) rectangle (axis cs:119.25,0.521739130434783);
\draw[draw=color0,fill=color0] (axis cs:119.75,0) rectangle (axis cs:120.25,0.478260869565217);
\draw[draw=color0,fill=color0] (axis cs:120.75,0) rectangle (axis cs:121.25,0.478260869565217);
\draw[draw=color0,fill=color0] (axis cs:121.75,0) rectangle (axis cs:122.25,0.521739130434783);
\draw[draw=color0,fill=color0] (axis cs:122.75,0) rectangle (axis cs:123.25,0.478260869565217);
\draw[draw=color0,fill=color0] (axis cs:123.75,0) rectangle (axis cs:124.25,0.391304347826087);
\draw[draw=color0,fill=color0] (axis cs:124.75,0) rectangle (axis cs:125.25,0.608695652173913);
\draw[draw=color0,fill=color0] (axis cs:125.75,0) rectangle (axis cs:126.25,0.652173913043478);
\draw[draw=color0,fill=color0] (axis cs:126.75,0) rectangle (axis cs:127.25,0.391304347826087);
\end{axis}

\end{tikzpicture}

%% file: figs/results/msp/single_bits_num_imgs.tex
\begin{tikzpicture}

\begin{axis}[
	axis background/.style={BgStyle},
	axis line style={white},
	height=\axheight,
	legend cell align={left},
	legend style={draw opacity=1, draw=none, text opacity=1 , fill=white, fill opacity=0.6, font=\scriptsize},
	legend pos=south east,
	tick align=outside,
	tick pos=both,
	ticklabel style = {font=\scriptsize},
	width=\axwidth,
	x grid style={white},
	xmajorgrids,
	xmin=29, xmax=401,
	xtick style={color=white!33.3333333333333!black},
	y grid style={white},
	ymajorgrids,
	ymin=0.5, ymax=1.01,
	ytick style={color=white!33.3333333333333!black},
	xlabel={Number of images used for training},
	ylabel={Test accuracy},
	label style = {font=\footnotesize}
	]

\addplot [thick, color0]
table {%
	410 1
};
\addlegendentry{256$\times$256\,px}
\addplot [thick, color1]
table {%
	410 1
};
\addlegendentry{128$\times$128\,px}
\addplot [thick, color2]
table {%
	410 1
};
\addlegendentry{64$\times$64\,px}

\addplot [thick, black, mark=*, mark size=1, mark options={MarkStyle}]
table {%
	410 1
};
\addlegendentry{Bit 0}
\addplot [thick, black, mark=triangle*, mark size=1, mark options={MarkStyle,rotate=180}]
table {%
	410 1
};
\addlegendentry{Bit 1}
\addplot [thick, black, mark=square*, mark size=1, mark options={MarkStyle}]
table {%
	410 1
};
\addlegendentry{Bit 2}

\addplot [thick, color0, mark=*, mark size=1, mark options={MarkStyle,}]
table {%
400 1
300 0.984962406015038
200 0.944206008583691
150 0.802120141342756
100 0.762762762762763
75 0.665738161559889
50 0.66579634464752
40 0.597964376590331
30 0.568238213399504
};

\addplot [thick, color1, mark=*, mark size=1, mark options={MarkStyle,}]
table {%
400 1
300 1
200 0.982832618025751
150 0.932862190812721
100 0.846846846846847
75 0.782729805013928
50 0.673629242819843
40 0.732824427480916
30 0.575682382133995
};

\addplot [thick, color2, mark=*, mark size=1, mark options={MarkStyle,}]
table {%
400 1
300 1
200 1
150 1
100 0.996996996996997
75 0.991643454038997
50 0.95822454308094
40 0.926208651399491
30 0.799007444168734
};

\addplot [thick, color0, mark=triangle*, mark size=1, mark options={MarkStyle,rotate=180,}]
table {%
400 1
300 1
200 0.957081545064378
150 0.932862190812721
100 0.867867867867868
75 0.779944289693593
50 0.60313315926893
40 0.641221374045801
30 0.531017369727047
};

\addplot [thick, color1, mark=triangle*, mark size=1, mark options={MarkStyle,rotate=180,}]
table {%
400 1
300 1
200 1
150 0.996466431095406
100 0.948948948948949
75 0.919220055710306
50 0.804177545691906
40 0.641221374045801
30 0.640198511166253
};

\addplot [thick, color2, mark=triangle*, mark size=1, mark options={MarkStyle,rotate=180,}]
table {%
400 1
300 1
200 1
150 1
100 1
75 1
50 1
40 0.994910941475827
30 0.987593052109181
};

\addplot [thick, color0, mark=square*, mark size=1, mark options={MarkStyle,}]
table {%
400 0.96969696969697
300 0.992481203007519
200 0.948497854077253
150 0.876325088339223
100 0.684684684684685
75 0.576601671309192
50 0.678851174934726
40 0.508905852417303
30 0.570719602977668
};

\addplot [thick, color1, mark=square*, mark size=1, mark options={MarkStyle,}]
table {%
400 1
300 1
200 0.909871244635193
150 0.862190812720848
100 0.726726726726727
75 0.64066852367688
50 0.590078328981723
40 0.603053435114504
30 0.548387096774194
};

\addplot [thick, color2, mark=square*, mark size=1, mark options={MarkStyle,}]
table {%
400 1
300 1
200 1
150 1
100 1
75 0.997214484679666
50 0.945169712793734
40 0.954198473282443
30 0.858560794044665
};

\end{axis}

\end{tikzpicture}

%% file: figs/results/msp/single_bits_num_imgs_randomized.tex
\begin{tikzpicture}

\begin{axis}[
	axis background/.style={BgStyle},
	axis line style={white},
	height=\axheight,
	legend cell align={left},
	legend style={draw opacity=1, draw=none, text opacity=1 , fill=white, fill opacity=0.8, font=\scriptsize},
	legend pos=south west,
	tick align=outside,
	tick pos=both,
	ticklabel style = {font=\scriptsize},
	width=\axwidth,
	x grid style={white},
	xmajorgrids,
	xmin=99, xmax=801,
	xtick style={color=white!33.3333333333333!black},
	y grid style={white},
	ymajorgrids,
	ymin=0.5, ymax=1.01,
	ytick style={color=white!33.3333333333333!black},
	xlabel={Number of images used for training},
	ylabel={Test accuracy},
	label style = {font=\footnotesize}
	]

\addplot [thick, color0]
table {%
	810 1
};
\addlegendentry{256$\times$256\,px}
\addplot [thick, color1]
table {%
	810 1
};
\addlegendentry{128$\times$128\,px}
\addplot [thick, color2]
table {%
	810 1
};
\addlegendentry{64$\times$64\,px}

\addplot [thick, black, mark=*, mark size=1, mark options={MarkStyle}]
table {%
	810 1
};
\addlegendentry{Bit 0}
\addplot [thick, black, mark=triangle*, mark size=1, mark options={MarkStyle,rotate=180}]
table {%
	810 1
};
\addlegendentry{Bit 1}
\addplot [thick, black, mark=square*, mark size=1, mark options={MarkStyle}]
table {%
	810 1
};
\addlegendentry{Bit 2}

\addplot [thick, color0, mark=*, mark size=1, mark options={MarkStyle}]
table {%
	800 0.727272727272727
	700 0.59504132231405
	600 0.617117117117117
	500 0.601246105919003
	400 0.575829383886256
	300 0.583493282149712
	200 0.552334943639291
	100 0.538141470180305
};

\addplot [thick, color1, mark=*, mark size=1, mark options={MarkStyle}]
table {%
	800 0.727272727272727
	700 0.710743801652893
	600 0.743243243243243
	500 0.635514018691589
	400 0.649289099526066
	300 0.616122840690979
	200 0.589371980676328
	100 0.589459084604716
};

\addplot [thick, color2, mark=*, mark size=1, mark options={MarkStyle}]
table {%
	800 1
	700 0.991735537190083
	600 0.963963963963964
	500 0.968847352024922
	400 0.912322274881517
	300 0.850287907869482
	200 0.784219001610306
	100 0.672676837725381
};

\addplot [thick, color0, mark=triangle*, mark size=1, mark options={MarkStyle,rotate=180}]
table {%
	800 0.772727272727273
	700 0.818181818181818
	600 0.779279279279279
	500 0.744548286604361
	400 0.66824644549763
	300 0.669865642994242
	200 0.621578099838969
	100 0.585298196948682
};

\addplot [thick, color1, mark=triangle*, mark size=1, mark options={MarkStyle,rotate=180}]
table {%
	800 1
	700 0.909090909090909
	600 0.936936936936937
	500 0.878504672897196
	400 0.83175355450237
	300 0.773512476007677
	200 0.752012882447665
	100 0.646324549237171
};

\addplot [thick, color2, mark=triangle*, mark size=1, mark options={MarkStyle,rotate=180}]
table {%
	800 1
	700 1
	600 0.995495495495496
	500 1
	400 0.997630331753555
	300 0.986564299424184
	200 0.948470209339775
	100 0.877947295423024
};

\addplot [thick, color0, mark=square*, mark size=1, mark options={MarkStyle}]
table {%
	800 0.727272727272727
	700 0.661157024793388
	600 0.653153153153153
	500 0.607476635514019
	400 0.613744075829384
	300 0.571976967370441
	200 0.573268921095008
	100 0.556171983356449
};

\addplot [thick, color1, mark=square*, mark size=1, mark options={MarkStyle}]
table {%
	800 0.727272727272727
	700 0.661157024793388
	600 0.567567567567568
	500 0.613707165109034
	400 0.568720379146919
	300 0.54510556621881
	200 0.524959742351047
	100 0.528432732316227
};

\addplot [thick, color2, mark=square*, mark size=1, mark options={MarkStyle}]
table {%
	800 1
	700 0.975206611570248
	600 0.981981981981982
	500 0.934579439252336
	400 0.921800947867299
	300 0.879078694817658
	200 0.772946859903382
	100 0.687933425797503
};

\end{axis}

\end{tikzpicture}

%% file: figs/results/cyclone/single_bits_num_imgs.tex
\begin{tikzpicture}

\begin{axis}[
	axis background/.style={BgStyle},
	axis line style={white},
	height=\axheight,
	legend cell align={left},
	legend style={draw opacity=1, draw=none, text opacity=1 , fill=white, fill opacity=0.6, font=\scriptsize},
	legend pos=south east,
	tick align=outside,
	tick pos=both,
	ticklabel style = {font=\scriptsize},
	width=\axwidth,
	x grid style={white},
	xmajorgrids,
	xmin=19, xmax=301,
	xtick style={color=white!33.3333333333333!black},
	y grid style={white},
	ymajorgrids,
	ymin=0.75, ymax=1.01,
	ytick style={color=white!33.3333333333333!black},
	xlabel={Number of images used for training},
	ylabel={Test accuracy},
	label style = {font=\footnotesize}
	]

\addplot [thick, color0]
table {%
	810 1
};
\addlegendentry{Full image}
\addplot [thick, color1]
table {%
	810 1
};
\addlegendentry{140$\times$14\,px}
\addplot [thick, color2]
table {%
	810 1
};
\addlegendentry{80$\times$14\,px}

\addplot [thick, black, mark=*, mark size=1, mark options={solid, draw=black, thin}]
table {%
	810 1
};
\addlegendentry{Bit 0}
\addplot [thick, black, mark=triangle*, mark size=1, mark options={solid, draw=black, thin,rotate=180}]
table {%
	810 1
};
\addlegendentry{Bit 1}
\addplot [thick, black, mark=square*, mark size=1, mark options={solid, draw=black, thin}]
table {%
	810 1
};
\addlegendentry{Bit 2}

\addplot [thick, color0, mark=*, mark size=1, mark options={solid, draw=black, thin}]
table {%
	500 1
	400 1
	300 1
	200 1
	150 0.997613365155131
	100 0.965884861407249
	75 0.989858012170385
	50 0.945945945945946
	40 0.867424242424242
	30 0.855018587360595
	20 0.77007299270073
};

\addplot [thick, color1, mark=*, mark size=1, mark options={solid, draw=black, thin}]
table {%
	500 1
	400 1
	300 1
	200 1
	150 1
	100 1
	75 1
	50 0.994208494208494
	40 0.982954545454545
	30 0.938661710037175
	20 0.839416058394161
};

\addplot [thick, color2, mark=*, mark size=1, mark options={solid, draw=black, thin}]
table {%
	500 1
	400 1
	300 1
	200 1
	150 1
	100 1
	75 0.993914807302231
	50 0.996138996138996
	40 0.996212121212121
	30 0.979553903345725
	20 0.985401459854015
};

\addplot [thick, color0, mark=triangle*, mark size=1, mark options={solid, draw=black, thin,rotate=180}]
table {%
	500 1
	400 1
	300 1
	200 1
	150 1
	100 1
	75 0.997971602434077
	50 0.986486486486487
	40 0.948863636363636
	30 0.907063197026022
	20 0.821167883211679
};

\addplot [thick, color1, mark=triangle*, mark size=1, mark options={solid, draw=black, thin,rotate=180}]
table {%
	500 1
	400 1
	300 1
	200 1
	150 1
	100 1
	75 1
	50 1
	40 0.998106060606061
	30 0.99814126394052
	20 0.948905109489051
};

\addplot [thick, color2, mark=triangle*, mark size=1, mark options={solid, draw=black, thin,rotate=180}]
table {%
	500 1
	400 1
	300 1
	200 1
	150 1
	100 1
	75 1
	50 1
	40 1
	30 1
	20 0.990875912408759
};

\addplot [thick, color0, mark=square*, mark size=1, mark options={solid, draw=black, thin}]
table {%
	500 1
	400 1
	300 1
	200 1
	150 1
	100 0.997867803837953
	75 0.997971602434077
	50 0.994208494208494
	40 0.960227272727273
	30 0.901486988847584
	20 0.797445255474453
};

\addplot [thick, color1, mark=square*, mark size=1, mark options={solid, draw=black, thin}]
table {%
	500 1
	400 1
	300 1
	200 1
	150 1
	100 1
	75 1
	50 0.996138996138996
	40 0.996212121212121
	30 0.988847583643123
	20 0.932481751824818
};

\addplot [thick, color2, mark=square*, mark size=1, mark options={solid, draw=black, thin}]
table {%
	500 1
	400 1
	300 1
	200 1
	150 1
	100 1
	75 1
	50 1
	40 1
	30 1
	20 0.947080291970803
};

\end{axis}

\end{tikzpicture}